\documentclass[prd,preprintnumbers,showpacs,nofootinbib,superscriptaddress,notitlepage,floatfix]{revtex4-1}

\usepackage[T1]{fontenc}
\usepackage[utf8]{inputenc}
\usepackage{amsmath}
\usepackage{amsfonts}
\usepackage{amssymb}
\usepackage{physics}
\usepackage{slashed}
\usepackage{simplewick}
\usepackage{hyperref} 
\usepackage{mathrsfs}
\usepackage{subfigure}

\usepackage{siunitx}

\usepackage{graphicx}
\usepackage{float}

\usepackage{listings}

\usepackage{tikz}
\usetikzlibrary{calc}
\usetikzlibrary{patterns}
\usetikzlibrary{patterns}
\usetikzlibrary{positioning,decorations.pathmorphing,decorations.markings}
\usetikzlibrary{shapes,arrows}

\usepackage{feynmp-auto}
\usepackage{multirow}

\newcommand{\diag}{\ensuremath{\text{diag}}}

\DeclareRobustCommand{\eq}[1]{eq.~\eqref{eq:#1}}

\DeclareRobustCommand{\eqs}[2]{eqs.~\eqref{eq:#1} and \eqref{eq:#2}}

\DeclareRobustCommand{\fig}[1]{Fig.~\ref{fig:#1}}
\DeclareRobustCommand{\figs}[2]{Figs.~\ref{fig:#1} and \ref{fig:#2}}
\DeclareRobustCommand{\app}[1]{appendix~\ref{app:#1}}
\DeclareRobustCommand{\sec}[1]{Sec.~\ref{sec:#1}}

\DeclareRobustCommand{\eq}[1]{Eq.~(\ref{eq:#1})}
\DeclareRobustCommand{\eqs}[2]{Eqs.~(\ref{eq:#1}) and (\ref{eq:#2})}

\newcommand{\MS}{{\overline{\mathrm{MS}}}}

\newcommand{\eps}{\epsilon}

\newcommand{\nn}{\nonumber}

\newcommand{\myAffiliation}{Institute of Physics, National Yang Ming
  Chiao Tung University, Hsinchu 30010, Taiwan}

\usepackage{xcolor}

\begin{document}

\begin{abstract}
Parton distribution functions (PDFs) and light-cone distribution
amplitudes (LCDAs) are central non-perturbative objects of interest in
high-energy inelastic and elastic scattering, respectively. As a
result, an \textit{ab-initio} determination of these objects is highly
desirable.   In this paper we present theoretical details for
the calculation of the PDFs and LCDAs using a
heavy-quark operator product expansion method.  This strategy was
proposed in a previous paper [Phys. Rev. D 73, 014501 (2006)] for
computing higher moments of the PDFs using lattice QCD.  Its central
feature is the introduction of a fictitious, valence heavy quark.
In the current article, we show that the operator product expansion (OPE) of the hadronic matrix element we study can also be expressed as the convolution of a perturbative matching kernel and the corresponding light-cone distribution, which in principle can be inverted to determine the parton momentum fraction dependence. Regarding the extraction of higher moments, this work also provides the one-loop Wilson coefficients in the OPE formulas for the unpolarized PDF, helicity PDF and pseudo-scalar meson LCDAs. 
 Although these Wilson
coefficients for the PDFs can be inferred from existing results in the
literature, those for the LCDAs are new.
\end{abstract}

\preprint{MIT-CTP/5294}

\title{Parton physics from a heavy-quark operator product expansion: \\
Formalism and Wilson coefficients}

\author{William Detmold}
\email{wdetmold@mit.edu}
\affiliation{Center for Theoretical Physics, Massachusetts Institute
  of Technology,
Cambridge, MA 02139, USA}
\affiliation{The NSF AI Institute for Artificial Intelligence and Fundamental Interactions
}

\author{Anthony V. Grebe}
\email{agrebe@mit.edu}
\affiliation{Center for Theoretical Physics, Massachusetts Institute
  of Technology,
  Cambridge, MA 02139, USA}

\author{Issaku Kanamori}
\email{kanamori-i@riken.jp}
\affiliation{RIKEN Center for Computational Science, Kobe 650-0047, Japan}

\author{C.-J. David Lin}
\email{dlin@nycu.edu.tw}
\affiliation{\myAffiliation}
\affiliation{Centre for High Energy Physics, Chung-Yuan Christian University, Chung-Li, 32032, Taiwan}

\author{Robert J. Perry}
\email{perryrobertjames@gmail.com}
\affiliation{\myAffiliation}

\author{Yong Zhao}
\email{yong.zhao@anl.gov}
\affiliation{Physics Division, Argonne National Laboratory, Lemont, IL 60439, USA}
\affiliation{Physics Department, Brookhaven National Laboratory, Bldg. 510A, Upton, NY 11973, USA}

\collaboration{HOPE Collaboration}

\maketitle

\section{Introduction}

Factorization in Quantum Chromodynamics (QCD) allows one to describe many hard hadronic reactions as a convolution of a short-range (high-energy) perturbative kernel and a long-range (low-energy) process-independent function~\cite{Collins:1989gx}. 
Historically, the approach used to describe high-energy hadronic processes has been to calculate the short-range kernel using perturbation theory and to extract the long-distance piece from experimental data.
 The validity of this approach is established by the universality of the long-range contributions, which are intrinsic properties of the hadron. 

While the low-energy functions in QCD factorization can be measured from experimental data, they are also calculable within a non-perturbative field theoretic framework. Currently, the only non-perturbative first-principles approach to QCD is through lattice regularization, and such low-energy functions are prime candidates for a direct calculation using this framework. Recently the advances in computational power, numerical algorithms and theoretical understanding have enabled these quantities to be computed. The two simplest examples of these long-distance functions are the parton distribution functions (PDFs) introduced to explain deep inelastic scattering (DIS)~\cite{Bjorken:1968dy,Feynman:1969ej,Bjorken:1969ja}, and the light-cone distribution amplitudes (LCDAs) introduced to describe high-energy exclusive processes~\cite{Radyushkin:1977gp,Lepage:1980fj}.

Direct evaluations of the PDFs and LCDAs require the computation of matrix elements involving non-perturbative dynamics on the light-cone,  and hence are not possible in a Euclidean formulation of field theory  like lattice QCD.   Historically, this problem in lattice calculations has been addressed using Wilson's operator product expansion (OPE)~\cite{Wilson:1969zs,Wilson:1972ee} to express the non-local operator as an infinite sum of local operators.  Matrix elements of these local operators are  directly calculable within the framework of lattice QCD, and are related to moments of the PDFs and LCDAs.   In principle, knowledge of these moments facilitates the reconstruction of the PDFs and LCDAs.  See, for example, Refs~\cite{Kronfeld:1984zv,Martinelli:1987si,Best:1997qp} for earlier works employing this strategy.  Nevertheless, the renormalisation procedure for the local operators appearing in this approach requires subtractions of power divergences that arise from the breaking of the $O(4)$ Euclidean space-time symmetry.  As a result, the method is practically applicable only to the determination of a few leading moments.

Over the past two decades, various alternative strategies have been proposed to extract PDFs and LCDAs using lattice QCD~\cite{Aglietti:1998ur,Liu:1999ak,Detmold:2005gg,Braun:2007wv,Davoudi:2012ya,Ji:2013dva,Chambers:2017dov,Radyushkin:2017cyf,Ma:2017pxb}.  All of these strategies involve lattice computations for matrix elements of non-local operators. In this article, we expand the theoretical details of the proposal suggested in Ref.~\cite{Detmold:2005gg}.  This proposal is originally designed to allow access to higher moments with lattice QCD.  Its key ingredient is the introduction of a fictitious, valence heavy quark that enables additional control of an OPE for calculating higher moments.  Here we call this the heavy-quark operator product expansion (HOPE) method.  In the current paper we show how one may extend the HOPE method to enable the direct determination of $\xi$-dependence of the PDFs and LCDAs.  This is achieved, as explained in detail in Sec.~\ref{sec:Hope_qPDF}, by demonstrating that the hadronic amplitudes appearing in the HOPE approach can be expressed as the convolution of a perturbative matching kernel and either a PDF or LCDA.  Concerning the computation of the moments, we complement the discussions in Ref.~\cite{Detmold:2005gg} by analysing the convergence properties of the HOPE.   This analysis is presented in Sec.~\ref{sec:convg} below.  Furthermore, in this work we provide the relevant one-loop Wilson coefficients for extracting moments of the unpolarised and helicity PDFs, as well as the meson LCDAs.   This allows one to account for renormalisation scheme and scale dependence in the moments more precisely.  Although the Wilson coefficients for the PDFs can be inferred from results in Refs.~\cite{Gottschalk:1980rv,Gluck:1996ve} where the cross sections for heavy-quark production in charged-current DIS are computed,  we perform alternative calculations and describe them in Secs.~\ref{sec:nlopdf} and \ref{sec:helicity}.  Finally, in Sec.~\ref{sec:danlo}  we give one-loop results for the LCDAs, which are new.

\section{Definitions and Elementary Properties}

In order to make this paper self-contained, we give formal definitions for the PDF and LCDA, summarize their uses and motivate the necessity for computing these quantities from first principles.  Although we intend this calculation to be used for a determination of moments of the LCDA using Euclidean lattice field theory, in this work we proceed in Minkowski space and utilize the ``mostly negative'' metric: $g^{\mu\nu}=\diag(+1,-1,-1,-1)$. Results in Euclidean space can be obtained straightforwardly provided the analytic continuation of the matrix element considered is possible. Further discussion of this point is found in \sec{convg}.

\subsection{Parton Distribution Function}

Parton distribution functions were introduced in the context of the study of DIS~\cite{Bjorken:1968dy,Feynman:1969ej,Bjorken:1969ja} and are an important input in the Standard Model predictions of collisions at high-energy hadronic colliders like the LHC. Currently nucleon PDFs are extracted through global fits to a number of different cross sections~\cite{Harland-Lang:2014zoa,Ball:2017nwa,Alekhin:2017kpj,Hou:2019efy,Sato:2019yez}. However, it is a goal of lattice field theory to produce predictions of the PDFs from first-principle calculations in QCD of similar or better precision to those extracted from experiments. Formally the unpolarized PDF, $f_{i/H}(\xi,\mu^2)$ and helicity PDF, $\Delta f_{i/H}(\xi,\mu^2)$ for a quark of flavor $i$ carrying a fraction $\xi$ of momentum in a hadron $H$ are defined as
\begin{align}
\begin{split}
f_{i/H}(\xi,\mu^2)&=\int \frac{dx^-}{(4\pi)}\  e^{-i\xi P^+ x^-}\bra{H(\mathbf{p},\mathbf{s})}\overline{\psi}_i(x^-)\gamma^+ W[x^-,0]\psi_i(0)\ket{H(\mathbf{p},\mathbf{s})},
\end{split}
\\
\begin{split}
\Delta f_{i/H}(\xi,\mu^2)&=\frac{1}{s^+}\int \frac{dx^-}{(4\pi)}\  e^{-i\xi P^+ x^-}\bra{H(\mathbf{p},\mathbf{s})}\overline{\psi}_i(x^-)\gamma^+ \gamma_5 W[x^-,0]\psi_i(0)\ket{H(\mathbf{p},\mathbf{s})},
\end{split}
\end{align}
where $\mu^2$ is the renormalization scale of the light-cone separated operator, $s^\mu$ is the spin vector of the hadron of mass $m_H$, with $s^2=-m_H^2$ and $s\cdot p=0$, and
\begin{equation}
W[x,y]=\mathcal{P}\exp(ig\int_x^y dz_\mu\  A_a^\mu(z)t_a)
\end{equation}
is a Wilson line which ensures gauge invariance of the matrix element. We use light-cone coordinates where for a general four-vector $V^\mu$ the ``plus'' and ``minus'' components are defined as
\begin{align}
V^\pm&=\frac{1}{\sqrt{2}}(V^0\pm V^3)\,.
\end{align}
In order to keep the notation relatively simple, in what follows we shall drop explicit flavor labels unless they are necessary. Correspondingly, $f_{i/H}$ and $\Delta f_{i/H}$ will be denoted as $f_{H}$ and $\Delta f_{H}$, respectively.

\subsection{Light-cone Distribution Amplitude}

The LCDA was originally introduced in order to study the asymptotically large momentum transfer limit of the pion electromagnetic form factor~\cite{Radyushkin:1977gp}. At around the same time, it was realized~\cite{Lepage:1980fj} that meson and baryon LCDAs were the central objects of interest in describing a number of exclusive process at high energy. In light-cone gauge where $A^+=0$, the LCDA $\phi_M(\xi,\mu^2)$ of a meson $M$ is interpreted as the probability amplitude for converting the meson into a collinear quark-antiquark pair with longitudinal momentum fractions $(1+\xi)/2$ and $(1-\xi)/2$, respectively, at the renormalization scale $\mu^2$. For pseudoscalar mesons, the LCDA is defined by the matrix element~\cite{Radyushkin:1977gp,Lepage:1980fj}
\begin{align}
\bra{\Omega}\overline{\psi}(z)\gamma_\mu \gamma_5 W[z,-z]\psi(-z)\ket{M(\mathbf{p})} = if_M p_\mu\int_{-1}^1 d\xi\  e^{-i\xi p\cdot z }\phi_M(\xi,\mu^2)\,,
\label{eq:lcda}
\end{align}
where $\bra{\Omega}$ is the physical vacuum state, $z^2=0$ is a light-like separation and $f_M$ is the pseudoscalar meson decay constant defined by the $z\to0$ limit of Eq.~\eqref{eq:lcda}. The most studied of the LCDAs is that of the pion. In the isospin limit where the masses of the up and down quarks are degenerate, the pion LCDA is symmetric under the interchange $\xi\to -\xi$, that is
\begin{equation}
\phi_\pi(\xi,\mu^2)=\phi_\pi(-\xi,\mu^2).
\end{equation}
We shall assume isospin symmetry in this work. At leading logarithmic accuracy, the natural orthogonal polynomial basis for the LCDA is the set of Gegenbauer polynomials~\cite{Lepage:1980fj,Efremov:1979qk},
\begin{equation}
\mathcal{C}=\{\mathcal{C}_n^{(3/2)}(\xi)\,|\,n=0,1,2\dots\}.
\end{equation}
The pion LCDA can thus be expressed as
\begin{equation}
\phi_\pi(\xi,\mu^2)=\frac{3}{4}(1-\xi^2)\sum_{n=0,2,\dots}^\infty \phi_n^\pi(\mu^2)\mathcal{C}_n^{(3/2)}(\xi).
\end{equation}
From this decomposition, it is clear that knowledge of the Gegenbauer moments $\phi_n^\pi(\mu^2)$ is sufficient to reconstruct the LCDA. In the limit that $\mu^2\to\infty$, $\phi_0(\mu^2)\to1$ and the higher moments go logarithmically to zero. As a result
\begin{equation}
\phi_\pi(\xi,\mu^2\to\infty)=\frac{3}{4}(1-\xi^2).\label{eq:lcda_asymp}
\end{equation}
The high-energy behavior of certain exclusive processes in QCD are controlled by this asymptotic form of the distribution amplitude given in Eq.~\eqref{eq:lcda_asymp}. In particular, in Ref.~\cite{Lepage:1980fj}, predictions are given for transition and electromagnetic form factors of the pion. The success of this formalism is particularly clear for the case of the $\gamma\gamma^*\to\pi^0$ transition form factor, where experimental data from Belle~\cite{Uehara:2012ag} appears to show the predicted asymptotic form, although previous experiments appeared to suggest that this prediction underestimated the data~\cite{Gronberg:1997fj,Aubert:2009mc}. For the case of the pion electromagnetic form factor, the situation is less clear. The most recent experimental data~\cite{Blok:2008jy,Huber:2008id} does not agree with the prediction of Ref.~\cite{Lepage:1980fj}. Although the extracted values of the pion form factor are in principle model dependent~\cite{Guidal:1997by,Vanderhaeghen:1997ts}, it has been shown that they are relatively insensitive to the model used~\cite{Perry:2018kok,Perry:2020hli}. Since the asymptotic form of the distribution amplitude is only expected to be valid at very high photon virtualities, it has been suggested that the discrepancy between theory and experiment can be traced to the use of Eq.~\eqref{eq:lcda_asymp} at energy scales where the LCDA has not yet evolved to its asymptotic form. 
While further high-energy measurements of the pion electromagnetic form factor are certainly required, it is clear that precise determinations of the pion LCDA are also necessary to fully understand this discrepancy.

\subsection{Current Theoretical Approaches}

Due to the importance of having accurate calculations of PDFs and LCDAs, much work has been done to obtain information about the these quantities using many different phenomenological approaches. As we have previously emphasized, the most reliable theoretical determinations of such non-perturbative objects are obtained from lattice QCD. The `traditional' approach involves a determination of the Mellin moments of either the PDF or LCDA. These moments may be written as linear combinations of the Gegenbauer moments. The Mellin moments of the unpolarized PDF $f$, helicity PDF $\Delta f$ and LCDA $\phi_M$ for meson $M$ are defined respectively as 
\begin{align}
a_{n,V}^{H}(\mu^2)&=
\int_0^1d\xi\ \xi^n  f_{H}(\xi,\mu^2) \,,\label{eq:anvH}
\\
a_{n,A}^{H}(\mu^2)&=
\int_0^1d\xi\ \xi^n \Delta f_{H}(\xi,\mu^2)\,,
\\
\expval{\xi^n}_{M}(\mu^2)&=\int_{-1}^1 d\xi\  \xi^n\phi_M(\xi,\mu^2).\label{eq:phiM}
\end{align}
With exact knowledge of all the Mellin moments, one is able to reconstruct the full $\xi$-dependence of the distributions.
In practice, numerical determinations restrict one to calculating a finite set of moments, and further difficulties appear when one attempts to compute matrix elements of twist-two operators with spin greater than four in a lattice regularized field theory~\cite{Gockeler:1996mu}. This is because the lattice regularization breaks the full rotation group $\text{SO}(4)$ and leads to mixing of operators with different mass dimensions, hence the appearance of power divergences. These power divergences make the determination of the higher moments difficult. Nevertheless, this approach has been well studied and has yielded results for the lowest three moments of the pion PDF~\cite{Best:1997qp,Guagnelli:2004ga,Capitani:2005jp,Abdel-Rehim:2015owa,Oehm:2018jvm,Alexandrou:2020gxs} (for the lattice calculations of nucleon PDF moments see a recent review in Ref.~\cite{Lin:2020rut}), as well as the first non-trivial moment of the pion LCDA and the first two moments of the kaon LCDA~\cite{Kronfeld:1984zv,Martinelli:1987si,Braun:2015axa,Oehm:2018jvm,Bali:2019dqc}

To go beyond the calculation of the lowest moments of the PDF or LCDA, several alternative approaches have been proposed. One method is based on the short-distance operator product expansion (OPE) of a current-current correlator~\cite{Braun:2007wv,Bali:2017gfr}, where the issue of power-divergent operator mixing can be avoided since the currents have a well-defined continuum limit, and one can in principle extract the $n\ge4$ moments of the LCDA given sufficient numerical precision. The current-current correlator has also been used in the calculation of PDFs in recent studies~\cite{Ma:2017pxb,Sufian:2020vzb}. 
A method has been proposed to restore the broken rotational symmetry~\cite{Davoudi:2012ya} which would reduce the power-divergent mixing so that the higher moments can be directly computed on the lattice. More recently, the formalism of large-momentum effective theory (LaMET)~\cite{Ji:2013dva,Ji:2014gla,Ji:2020ect} was proposed as a method to determine the $\xi$-dependence of the PDF~\cite{Alexandrou:2018pbm,Lin:2018pvv,Izubuchi:2019lyk}. In this approach, one calculates the ``quasi-PDF'' which is defined as the Fourier transform of equal-time quark bilinear corrrelators in a hadron state with large momentum. The LaMET approach has also been applied to the lattice calculation of the LCDAs for light mesons~\cite{Zhang:2017bzy,Chen:2017gck,Zhang:2020gaj}. The same matrix element that defines the quasi-PDF can also be used in a short-distance coordinate space expansion known as the ``Ioffe-time pseudo distribution'' approach~\cite{Radyushkin:2017cyf,Orginos:2017kos}. Using this formalism, several moments of the PDF have been determined, and the $\xi$-dependence has been determined via a fit to a parametrized {\it ansatz}~\cite{Joo:2019bzr,Gao:2020ito,Joo:2020spy}. A direct lattice calculation of the hadronic tensor has also been proposed and explored~\cite{Liu:1993cv,Liang:2019frk}. In a similar vein, an application of the Feynman-Hellman theorem allows indirect access to the hadronic tensor via a determination of two-point correlators coupled to external fields~\cite{Chambers:2017dov,Can:2020sxc}.

In this work we pursue a different approach: the HOPE method~\cite{Detmold:2005gg,Detmold:2018kwu,Detmold:2020lev}. This procedure was first proposed in Ref.~\cite{Detmold:2005gg} and builds on the conventional OPE by performing the numerical simulation of a current-current correlator with the currents containing a fictitious heavy quark species which leads to a number of advantages over the standard treatment~\cite{Detmold:2005gg}.  In the next section, we review this strategy and provide more details about its theoretical foundation and practical implementation.

\section{The HOPE Method}
\label{sec:the_hope}
The HOPE method was originally proposed in Ref.~\cite{Detmold:2005gg} as a way to extract information about the moments of PDFs and LCDAs via an analysis of matrix elements which we refer to as \textit{hadronic amplitudes}. We define these as
\begin{align}
T^{\mu\nu}(p,q)&=\int d^4z\  e^{iq\cdot z}\bra{H(\mathbf{p},\mathbf{s})}\mathcal{T}\{J_A^\mu(z/2)J_A^\nu(-z/2)\}\ket{H(\mathbf{p},\mathbf{s})},\label{eq:compton_amplitude}
\\
V^{\mu\nu}(p,q)&=\int d^4z\ e^{iq\cdot z}\bra{\Omega}\mathcal{T}\{J_A^\mu(z/2)J_A^\nu(-z/2)\}\ket{M(\mathbf{p})},\label{eq:lcda_amplitude}
\end{align}
where $\mathcal{T}$ is the time-ordering symbol and the heavy-light axial-vector current $J_A^\mu(z)$ is given by
\begin{equation}
J_A^\mu(z)=\overline{\Psi}(z)\gamma^\mu \gamma_5 \psi(z)+\overline{\psi}(z)\gamma^\mu \gamma_5 \Psi(z).\label{eq:hope_current}
\end{equation}
We denote the fictitious heavy quark species as $\Psi$ and the light quark species as $\psi$. Note that other choices for the Dirac matrix structure in the current are also possible. The application of the HOPE method to $T^{\mu\nu}(p,q)$ enables one to extract the moments of the unpolarized and helicity PDFs. The matrix element $V^{\mu\nu}(p,q)$ allows a determination of the LCDA moments since both the current-current operator shown above and the operator in Eq.~\eqref{eq:lcda} can be expanded in terms of the same local operators~\cite{Brodsky:1980ny}.

The HOPE strategy is characterized by its use of a fictitious flavor-changing heavy-light current (Eq.~\eqref{eq:hope_current}). The heavy quark is quenched and as a result only propagates between the two currents~\cite{Detmold:2005gg}. We note that the use of a quenched heavy quark removes certain higher-twist diagrams entirely from the calculation and also provides a source of higher-twist suppression beyond the large momentum transfer at the currents due to the the heavy quark mass, $m_\Psi$. In order for the heavy quark mass to be considered a `large' scale we require that
\begin{equation}
\Lambda_\text{QCD}\ll m_\Psi\sim \sqrt{Q^2},
\end{equation}
where $Q^2=-q^2$. 

To apply the HOPE method, we define the operator
\begin{equation}
t^{\mu\nu}(q;m_\Psi)=\int d^4z\ e^{iq\cdot z}\mathcal{T}\{J_A^\mu(z/2)J_A^\nu(-z/2)\}.
\end{equation}
and perform an OPE. In the case of the symmetrized isovector operator, this yields~\cite{Detmold:2005gg}
%
\begin{align}
t^{\{\mu\nu\}}(q;m_\Psi)&=\frac{4i}{\tilde{Q}^2} \sum_{\substack{n=0\\ \text{even}}}^\infty \frac{(2q_{\mu_1})\dots (2q_{\mu_n})}{\tilde{Q}^{2n}}C_{1,n}(\tilde{Q}^2,m_\Psi^2,\mu^2)\mathcal{O}_{n+2,V}^{\mu\nu \mu_1\dots\mu_{n}}(\mu)
\nn\\
&\qquad +i g^{\mu\nu}\sum_{\substack{n=2\\ \text{even}}}^\infty\frac{(2q_{\mu_1})\dots (2q_{\mu_n})}{\tilde{Q}^{2n}}C_{2,n}(\tilde{Q}^2,m_\Psi^2,\mu^2)\mathcal{O}_{n,V}^{\mu_1\dots\mu_{n}}(\mu)\nn\\
&\qquad -{2\over \tilde Q^2}g^{\mu\nu}m_\Psi \sum_{\substack{n=0\\ \text{even}}}^\infty\frac{(2q_{\mu_1})\dots (2q_{\mu_n})}{\tilde{Q}^{2n}}C_{3,n}(\tilde{Q}^2,m_\Psi^2,\mu^2)\hat{\mathcal{O}}_{n}^{\mu_1\dots\mu_{n}}(\mu) \nn\\
&\qquad  -{4i\over \tilde Q^2}q^{\{\mu }\sum_{\substack{n=1\\ \text{odd}}}^\infty\frac{(2q_{\mu_1})\dots (2q_{\mu_n})}{\tilde{Q}^{2n}}C_{4,n}(\tilde{Q}^2,m_\Psi^2,\mu^2)\mathcal{O}_{n+1,V}^{\nu\}\mu_1\dots\mu_{n}} +\text{higher twist}(\mu)
,\label{eq:hope_compton}
\end{align}
%
where $C_{i,n}$ are the Wilson coefficients and we have defined
\begin{equation}
\tilde{Q}^2=-q^2+m_\Psi^2.\label{eq:hope_twist_suppression}
\end{equation}
Higher twist contributions are supressed by at least one extra inverse power of $\tilde{Q}$. The operators occurring on the right hand side of Eq.~\eqref{eq:hope_compton} are the well-known standard twist-two operators:
\begin{align}
\mathcal{O}_{n,V}^{\mu_1\dots \mu_n}&=\overline{\psi}\gamma^{\{\mu_1}(i\overset{\leftrightarrow}{D}\vphantom{D}^{\,\mu_2})\dots(\overset{\leftrightarrow}{D}\vphantom{D}^{\,\mu_n\}})\psi-\tr,\label{eq:ov}
\\
\mathcal{O}_{n,A}^{\mu_1\dots \mu_n}&=\overline{\psi}\gamma^{\{\mu_1}\gamma_5(i\overset{\leftrightarrow}{D}\vphantom{D}^{\,\mu_2})\dots(i\overset{\leftrightarrow}{D}\vphantom{D}^{\,\mu_n\}})\psi-\tr,\label{eq:oa}
\end{align}
and the twist-three operators:
\begin{equation}
\hat{\mathcal{O}}_{n}^{\mu_1\dots \mu_n}=\overline{\psi}(i\overset{\leftrightarrow}{D}\vphantom{D}^{\,\{\mu_1})(i\overset{\leftrightarrow}{D}\vphantom{D}^{\,\mu_2})\dots(i\overset{\leftrightarrow}{D}\vphantom{D}^{\,\mu_n\}})\psi-\tr,
\end{equation}
where $i$ is the quark flavor, $\{\cdots\}$ denotes total symmetrization of the Lorentz indices and trace subtraction, and
\begin{equation}
\overset{\leftrightarrow}{D}\vphantom{D}^{\,\mu}=\frac{1}{2}(\overset{\rightarrow}{D}\vphantom{D}^{\,\mu}-\overset{\leftarrow}{D}\vphantom{D}^{\,\mu}).
\end{equation}
Lorentz covariance allows us to write the general form for the twist-two matrix elements as 
\begin{align}
\bra{H(\mathbf{p},\mathbf{s})}\mathcal{O}_{n,V}^{\mu_1\dots \mu_n}\ket{H(\mathbf{p},\mathbf{s})}&=2a_{n,V}^{H}(\mu^2)[p^{\mu_1}\dots p^{\mu_n}-\tr],\label{eq:localops_vector}
\\
\bra{H(\mathbf{p,s})}\mathcal{O}_{n,A}^{\mu_1\dots \mu_n}\ket{H(\mathbf{p,s})}&=2a_{n,A}^{H}(\mu^2)[s^{\{\mu_1}p^{\mu_2}\dots p^{\mu_n\}}-\tr],
\\
\bra{H(\mathbf{p},\mathbf{s})}\hat{\mathcal{O}}_{n}^{\mu_1\dots \mu_n}\ket{H(\mathbf{p},\mathbf{s})}&=2b_{n}^{H}(\mu^2)[p^{\mu_1}\dots p^{\mu_n}-\tr],
\\
\bra{\Omega}\mathcal{O}_{n,A}^{\mu_1\dots \mu_n}\ket{M(\mathbf{p})}&=f_M\expval{\xi^{n-1}}_M(\mu^2)[p^{\mu_1}p^{\mu_2}\dots p^{\mu_n}-\tr],
\end{align}
where $\ket{H(\mathbf{p},\mathbf{s})}$ is a general hadronic state $H$ with mometum $\mathbf{p}$ and spin $\mathbf{s}$, $\ket{M(\mathbf{p})}$ is a general pseudoscalar mesonic state with momentum $\mathbf{p}$, $a_{n,V}^{H}$, $a_{n,A}^{H}$, $b_{n}^{H}$ and $\expval{\xi^{n}}_M$ are the Mellin moments, and $f_M$ is the meson decay constant. 

Explicitly, for the symmetric case of Eq.~\eqref{eq:compton_amplitude} in a pion, we have
\begin{align}\label{eq:hope_compton2}
T^{\{\mu\nu\}}(p,q)&=\bra{\pi(\mathbf{p})}t^{\{\mu\nu\}}(q;m_\Psi)\ket{\pi(\mathbf{p})}
\\
&=\frac{i}{\tilde{Q}^2}\bigg(4p^\mu p^\nu\sum_{\substack{n=0\\ \text{even}}}^\infty \tilde{\omega}^n {C}_{1,n}(\tilde{Q}^2,m_\Psi^2,\mu^2)2a_{n+2,V}^{\pi}(\mu^2)
+\tilde{Q}^2 g^{\mu\nu}\sum_{\substack{n=2\\ \text{even}}}^\infty\tilde{\omega}^n{C}_{2,n}(\tilde{Q}^2,m_\Psi^2,\mu^2)2a^{\pi}_{n,V}(\mu^2)\nn
\\
&+2ig^{\mu\nu}m_\Psi\sum_{\substack{n=0\\ \text{even}}}^\infty\tilde{\omega}^n{C}_{3,n}(\tilde{Q}^2,m_\Psi^2,\mu^2)2b_n^{\pi}(\mu^2)-2(p^\mu q^\nu+q^\mu p^\nu) \sum_{\substack{n=1\\ \text{odd}}}^\infty\tilde{\omega}^n{C}_{4,n}(\tilde{Q}^2,m_\Psi^2,\mu^2)2a_{n+1,V}^{\pi}(\mu^2)
\bigg)\nn,
\end{align}
where our expansion parameter is
\begin{equation}
\tilde{\omega}=\frac{1}{\tilde{x}_B}=\frac{2p\cdot q}{\tilde{Q}^2},\label{eq:hope_expansion_parameter}
\end{equation}
and $p^\mu$ is the momentum of the external hadronic state. Target mass corrections can be resummed using the standard techniques~\cite{Nachtmann:1973mr,Georgi:1976vf,Georgi:1976ve,Wandzura:1977ce}, and the explicitly resummed expression for \eq{hope_compton2} can be found in Ref.~\cite{Detmold:2005gg}. In either case, the application of the formalism to lattice QCD is as follows: first, numerically calculate the required hadronic amplitude, then fit the HOPE expression to lattice data with the moments as free parameters. Note that the applicability of the approach requires $|\tilde{\omega}| < 1$, leading to kinematical suppression of higher moments. Successful application of the HOPE method requires $\tilde{Q}^2\gg\Lambda_\text{QCD}^2$, so in order to enhance the contributions from higher moments for fixed $q^\mu$, one must increase the hadron momentum $|\mathbf{p}|$. This point has been discussed previously in Ref.~\cite{Detmold:2020lev}. 

\subsection{Relationship to Other Approaches} 
\label{sec:Hope_qPDF}

The HOPE method allows extraction of information about light-cone quantities from a current-current correlator where the standard current operator is replaced by that given in Eq.~\eqref{eq:hope_current}. While the use of this fictitious flavor changing current implies that the resulting matrix elements are not observables of standard QCD, the quenching of the heavy quark ensures the modifications may be wholly absorbed into the Wilson coefficients and thus the resulting moments correspond to those of QCD. As a result it is possible to show that the amplitude may be written in a factorized form as is done in the quasi-PDF~\cite{Ji:2013dva,Ji:2014gla,Ji:2020ect}, ``lattice cross section''~\cite{Ma:2017pxb} (LCS) or pseudo-PDF~\cite{Radyushkin:2017cyf,Orginos:2017kos} approaches. To do this, we start from Eq.~\eqref{eq:compton_amplitude} and use the HOPE expression for the operator given in Eq.~\eqref{eq:hope_compton}. By taking $\mu=3$, $\nu=0$, we can remove contributions from the twist-3 operators and as a result the HOPE expression for the pion hadronic amplitude $T^{\mu\nu}(p,q)$ is~\cite{Detmold:2005gg}:
\begin{align}
T^{\{30\}}(p,q)&=8i\frac{p^3p^0}{\tilde{Q}^2}\sum_{n=0,\text{even}}^\infty \tilde{\omega}^n {C}_{1,n}(\tilde{Q}^2,m_\Psi^2,\mu^2) a_{n+2,V}^{\pi}(\mu^2)-4i\frac{p^0q^3+p^3q^0}{\tilde{Q}^2}\sum_{n=1,\text{odd}}\tilde{\omega}^n {C}_{4,n}(\tilde{Q}^2,m_\Psi^2,\mu^2)a_{n+1,V}^{\pi}(\mu^2)\nn
\\
&\qquad+\text{higher-twist}.
\end{align}
Importantly, this relation is valid to twist-two accuracy, with higher-twist contributions being suppressed by at least an extra factor of $\Lambda_{\text{QCD}}/\tilde{Q}$.
When the momentum of the hadron becomes large, it also serves to suppress the target-mass effects which are of higher twist. To see this, let us compare the twist-two operator ${\cal O}_{n,V}^{\mu_0\mu_1}$ to the twist-four counter-part $g^{\mu_0\mu_1}{\cal O}_{n,V}^{(4)}$, which is given by
\begin{equation}
{\cal O}^{(4)}_{n,V}=\overline{\psi} (i\overset{\leftrightarrow}{\slashed{D}})\psi.
\end{equation}
Their contributions to the OPE are
\begin{align}
	(2q_{\mu_0})(2q_{\mu_1})\bra{H(\mathbf{p},\mathbf{s})}\mathcal{O}_{n,V}^{\mu_0\mu_1}\ket{H(\mathbf{p},\mathbf{s})} &= 2a_{n,V}^{H} \left[(2q\cdot p) ^2 - q^2 m_H^2\right]\,,\\
	(2q_{\mu_0})(2q_{\mu_1})g^{\mu_0\mu_1}\bra{H(\mathbf{p},\mathbf{s})}\mathcal{O}^{(4)}_{n,V}\ket{H(\mathbf{p},\mathbf{s})} &= 2a_{n,V}^{(4)}[4q^2 m_H^2]\,.
\end{align}
where $a_{n,V}^{(4)}$ is the Mellin moment for the twist-four operator ${\cal O}^{(4)}_{n,V}$. Thus we see that for kinematics where $q\cdot p \sim -q^2 \gg m_H^2$, both the target mass corrections and twist-four corrections are suppressed.

Using Eq.~\eqref{eq:anvH}, we may write
\begin{align}
T^{30}(p,q)&=\int_{0}^1 d\xi\ f_{H}(\xi,\mu^2) K(\xi,\tilde{\omega},\tilde{Q}^2,\mu^2)+\text{higher-twist},\label{eq:compton_factorized}
\end{align}
where the matching kernel is
\begin{align}\label{eq:kernel}
K(\xi,\tilde{\omega},\tilde{Q}^2,&\mu^2)=\sum_{n=0,\text{even}}^\infty\bigg[8i\frac{p^3p^0}{\tilde{Q}^2} \xi^2(\xi\tilde{\omega})^n {C}_{1,n}(\tilde{Q}^2,m_\Psi^2,\mu^2)-4i\frac{p^0q^3+p^3q^0}{\tilde{Q}^2}\xi(\xi\tilde{\omega})^{n+1} {C}_{4,n+1}(\tilde{Q}^2,m_\Psi^2,\mu^2)\bigg],
\end{align}
and we again emphasize that the flavor labels have been suppressed in the above definitions of the PDF and matching kernel.
Assuming $|\tilde{\omega}|<1$, we can re-sum the resulting geometric series to find a closed-form expression for the matching kernel~\cite{Martinelli:1998hz,Chambers:2017dov}. More generally, one may express the amplitude as the convolution of the PDF and the short-range kernel $K$. Importantly, one must ensure that $\tilde{Q}$ is large enough to isolate the twist-two contribution, and so that the matching coefficients can be calculated using perturbation theory.

According to Eq.~\eqref{eq:compton_factorized}, the hadronic amplitude defined by Eq.~\eqref{eq:compton_amplitude} may also be analyzed using a factorization theorem similar to the quasi-PDF, LCS or pseudo-PDF approaches. If we obtain the hadron amplitude as a continuous function in $\tilde\omega$ in its full range, then in principle we can invert \eq{compton_factorized} as is done in the quasi-PDF approach. Note that the kinematics we require to ensure a valid OPE is equivalent to that for the factorization of the amplitude into the light-cone quantity of interest and a perturbatively calculable matching kernel. To obtain large $\tilde \omega \sim 1$, we need large pion momentum. With a limited range of $\tilde \omega$, it is difficult to perform the inversion, so we have to rely on either a model fit of the PDF to the hadronic amplitude data using \eq{compton_factorized}, or we can fit the lowest moments using the OPE formula. In this regard, both strategies allow us to extract the same amount of information about the PDF, but the latter is model independent because the it does not introduce uncontrollable systematic effects from modeling the PDF to perform the inversion.

Generically, fitting lattice data to OPE formulae allows for the extraction of higher moments of the PDFs and LCDAs. 
  This evades the need for the direct evaluation of matrix elements of local operators, which involve the subtraction of power divergences in the renormalisation procedure.  In principle, having information for all moments enables one to reconstruct the PDFs and LCDAs.   Indeed, as demonstrated in the above section, the HOPE hadronic amplitudes allow us to determine directly the $\xi$-dependence 
  in these latter quantities, since the matching kernel, $K$, in Eqs.~(\ref{eq:compton_factorized}) and (\ref{eq:kernel}) is perturbatively computable.   The same general idea has been realised in different practical ways.  For example, similar proposals for obtaining information about parton physics can also be found in Refs.~\cite{Braun:2003rp,Chambers:2017dov}.   Compared to these other proposals, the introduction of the fictitious (valence) heavy quark in the
  HOPE strategy provides several advantages \footnote{The presence of an extra large mass scale modifies the Wilson coefficients $C_{j,n} (j=1,2,3,4)$ that have been derived in the massless case~\cite{Gross:1973ju,Gross:1974cs,Georgi:1951sr,Calvo:1977ba,DeRujula:1976baf,Altarelli:1978id,Kodaira:1978sh,Kodaira:1979ib,Kodaira:1979pa}. Therefore, a precise determination of the PDF or LCDA would necessitate the recalculation of $C_{j,n}$ and $K$ with the explicit dependence on the heavy quark mass. One can determine $K$ from the the Wilson coefficients $C_{j,n}$ presented in~\sec{oneloop}.}.  As already explained in Ref.~\cite{Detmold:2005gg}, the heavy quark provides a large scale that offers an additional parameter to control the OPE, and its valence nature is crucial for suppressing higher-twist effects.   Here we would like to point out two other assets that arise from the introduction of this heavy quark.  First, in the setting where $m_{\Psi}^{2}\gtrsim (p+q)^{2}$, it is natural that no on-shell state can
 appear between the two currents in the Minkowski-space  counterpart of the HOPE hadronic amplitudes.  This means that their analytic continuation is straightforward.  
 Secondly, the presence of the heavy quark serves to suppress the matrix element at large distances, and thus ensure that the effect of truncating the finite numerical data when performing the temporal Fourier transform is negligible.

\subsection{Convergence and Analytic Structure of HOPE Hadronic Amplitude}
\label{sec:convg}

In this section, we shall analyze the convergence properties of the
HOPE hadronic amplitude $T^{\mu\nu}(p,q)$. A similar analysis can also
be made for the matrix element $V^{\mu\nu}(p,q)$.
 The materials presented in this section are
  important for examining the validity of the HOPE method, and they
  complement the discussion in Ref.~\cite{Detmold:2005gg} where the
  approach was first proposed.
To begin with, we note that $T^{\mu\nu}(p,q)$ is a rank-2 tensor and thus the most general Lorentz covariant decomposition is
\begin{align}
T^{\mu\nu}(p,q)&= g^{\mu\nu}A_1(q^2,p\cdot q;m_\Psi)+p^\mu p^\nu A_2(q^2,p\cdot q;m_\Psi)+p^\mu q^\nu A_3(q^2,p\cdot q;m_\Psi)+q^\mu p^\nu A_4(q^2,p\cdot q;m_\Psi)\nn
\\
&\qquad+q^\mu q^\nu A_5(q^2,p\cdot q;m_\Psi)
+i\epsilon^{\mu\nu\alpha\beta}q_\alpha p_\beta A_6(q^2,p\cdot q;m_\Psi),
\end{align}
where the $A_i$ are scalar functions which are dependent on all possible kinematic variables. In the HOPE method, one performs an OPE as described in Ref~\cite{Detmold:2005gg}. We may equivalently express the scalar functions $A_i$ in terms of the variables 
\begin{equation}
\tau=Q^2/\tilde{Q}^2,\label{eq:tau}
\end{equation}
$\tilde{Q}^2$ and $\tilde{\omega}$ (defined in Eq.~\eqref{eq:hope_twist_suppression} and Eq.~\eqref{eq:hope_expansion_parameter}, respectively) as
\begin{equation}
A_i=A_i(\tilde{Q}^2,\tilde{\omega},\tau).
\end{equation}
Understanding the analytic properties of these scalar functions will enable us to understand the radius of convergence for the HOPE strategy in terms of the expansion variable $\tilde{\omega}$. Since the intention of this work is to describe the theoretical background for a numerical calculation using lattice QCD, we note that the relevant Euclidean constraint can be found by taking $q_0\to iq_4$ in $\tilde{\omega}$, provided the amplitude is analytic for this value of $q_0$. Since $\tilde{\omega}$ is in general complex, we seek to determine the radius of convergence in the complex plane. To study this, we must analyze the general analytical properties of the amplitude without recourse to perturbation theory. We shall consider the amplitude purely in Minkowski space and derive relations for the physical poles and branch points. Information regarding the locations of these singularities is important for determining the radius of convergence. 

\begin{figure}
\centering
\includegraphics[scale=1,trim={5cm 18cm 8cm 2cm},clip]{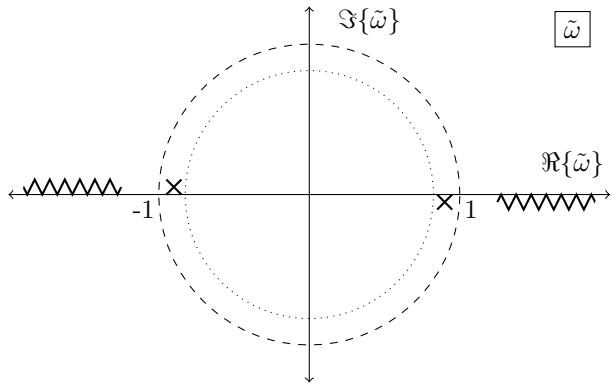}
\caption{The analytic structure of HOPE Compton amplitude. The radius of convergence is determined by the location of the nearest singularity, which is represented as a cross. The first branch point appears at $|\tilde{\omega}_\text{b.p.}|=1+[(m_\text{HL}+m_\pi)^2-m_\Psi^2-m_\pi^2]/\tilde{Q}^2$ and the branch cut is represented by the zigzag.
  In order to ensure a convergent expression one should choose a value of $|\tilde{\omega}|<\tilde{\omega}_\text{pole}=1-(m_\pi^2+m_\Psi^2-m_\text{HL}^2)/\tilde{Q}^2$.}
\label{fig:analytic_structure}
\end{figure}

To discuss the analytic properties of the scalar functions $A_i(\tilde{Q}^2,\tilde{\omega},\tau)$, it is useful to express these amplitudes in terms of standard Mandelstam variables. Noting that the matrix element we consider is in the forward limit, we write
\begin{align}
s&=(p+q)^2\,,\ \ \ \ \ t=0\,,\ \ \ \ \  u=(p-q)^2\,.
\end{align}
Conservation of momentum allows us to relate $u$ to $q^2$ and $s$. We may express $s$ in terms of the HOPE variables:
\begin{align}
s&=m_\pi^2+\tau \tilde{Q}^2(\tau^{-1}\tilde{\omega}-1) =m_\pi^2+\tilde{Q}^2(\tilde{\omega}-\tau)\,.
\end{align}
The lowest pole $\tilde{\omega}_\text{pole}$ in $A_i(\tilde{Q}^2,\tilde{\omega},\tau)$ appears at a value of $s=m_\text{HL}^2$ corresponding to the propagation of a heavy-light axial-vector meson with mass $m_{\rm HL}$ between the two inserted currents: 
\begin{align}
\tilde{\omega}_\text{pole}&=\tau+\frac{m_\text{HL}^2-m_\pi^2}{\tilde{Q}^2} =1-\frac{m_\pi^2+m_\Psi^2-m_\text{HL}^2}{\tilde{Q}^2}\,.
\end{align}
The first hadronic branch point which will appear is due to the two-body threshold ($s=(m_\pi+m_\text{HL})^2$). The corresponding position of the branch point in terms of $\tilde{\omega}$ is
\begin{align}
\tilde{\omega}_\text{b.p.}&=\tau+\frac{(m_\text{HL}+m_\pi)^2-m_\pi^2}{\tilde{Q}^2} =1+\frac{(m_\text{HL}+m_\pi)^2-m_\Psi^2-m_\pi^2}{\tilde{Q}^2}\,.
\end{align}
Further branch points appear due to the opening of more channels as the energy increases. Finally, from Eq.~\eqref{eq:compton_amplitude} it is possible to see that the amplitude is invariant under the combined interchange
\begin{equation}
q\leftrightarrow -q,\, \ \ \ \ \mu\leftrightarrow\nu.
\end{equation}
Under this interchange the roles of $s$ and $u$ are reversed. Above threshold this corresponds to the physical region of the $u$-channel process. In terms of $\tilde{\omega}$, this leads to
\begin{equation}
\tilde{\omega}\leftrightarrow -\tilde{\omega},
\end{equation}
and thus for each singularity or branch point we have discussed in $\tilde{\omega}$ due to a physical process in the $s$-channel there will be a corresponding singularity or branch point at $-\tilde{\omega}$ for the $u$-channel process. The resulting analytic structure of the amplitude is shown in Fig.~\ref{fig:analytic_structure}.
From this discussion of the analytic properties of the hadronic amplitude one may be assured that the HOPE method converges when performing numerical simulations provided one chooses kinematics such that 
\begin{equation}
|\tilde{\omega}|<\tilde{\omega}_\text{pole}=1-\frac{m_\pi^2+m_\Psi^2-m_\text{HL}^2}{\tilde{Q}^2}.
\end{equation}
Note that in the limit that the heavy quark mass becomes infinite, $\tilde{\omega}_\text{pole}\to1$ and $\tilde{\omega}_\text{b.p.}\to1$.

\hspace{1cm}

\section{One-Loop Wilson Coefficients for PDF and LCDA}
\label{sec:oneloop}

In this section, we derive the one-loop Wilson coefficients in the HOPE of the hadronic amplitude in \eqs{compton_amplitude}{lcda_amplitude} for both the PDF and pseudoscalar meson LCDA cases. Compared to the OPE for the hadronic amplitude from light-quark current-current correlators which has been previously studied~\cite{Gross:1973ju,Gross:1974cs,Georgi:1951sr,Calvo:1977ba,DeRujula:1976baf,Altarelli:1978id,Bardeen:1978yd,Kodaira:1978sh,Kodaira:1979ib,Kodaira:1979pa,Braaten:1982yp,Mueller:1997ak,Braun:2003rp,Braun:2007wv}, the heavy-quark mass in the HOPE method modifies the Wilson coefficients. Note that for the unpolarized and helicity PDF cases, the Wilson coefficents can be extracted from the cross section for heavy quark production through the charged current in neutrino DIS~\cite{Gottschalk:1980rv,Gluck:1996ve}, while for the LCDA case there is no known results in the literature. Therefore, our calculation of the latter is new in this work.

\begin{figure}
	\centering
\subfigure[]{\label{fig:tree}
	\includegraphics[width=0.25\linewidth]{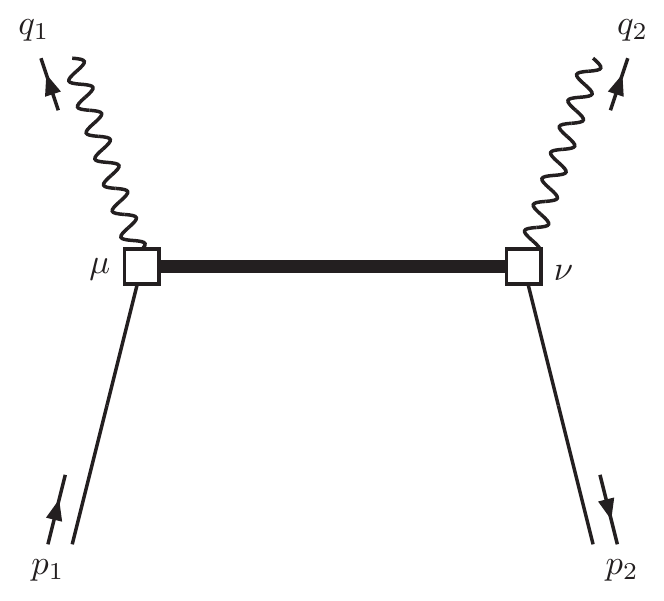}
}
\subfigure[]{\label{fig:box}
	\includegraphics[width=0.25\linewidth]{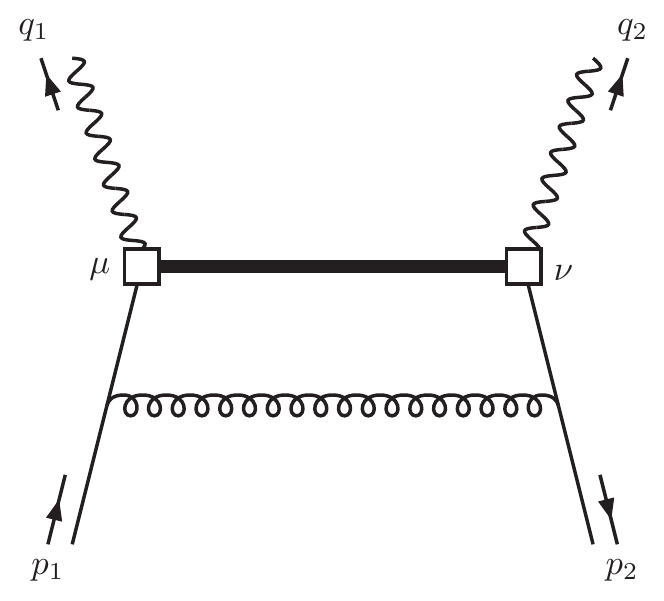}
}
\subfigure[]{\label{fig:v1}
	\includegraphics[width=0.25\linewidth]{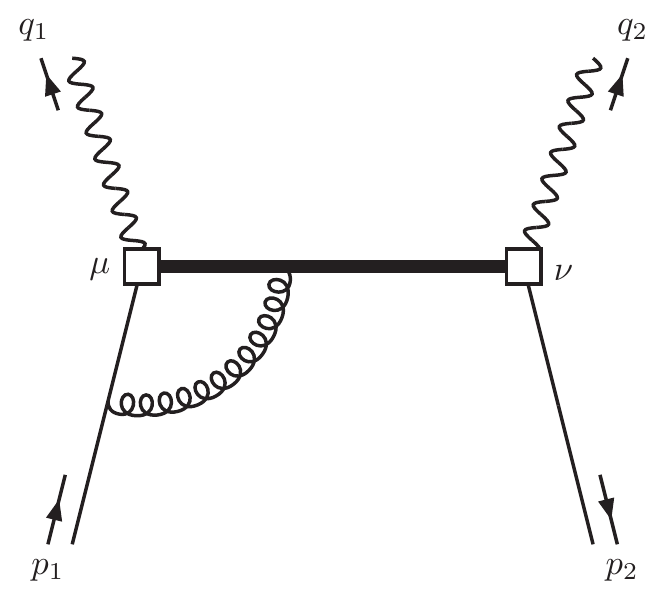}
}
\subfigure[]{\label{fig:v2}
	\includegraphics[width=0.25\linewidth]{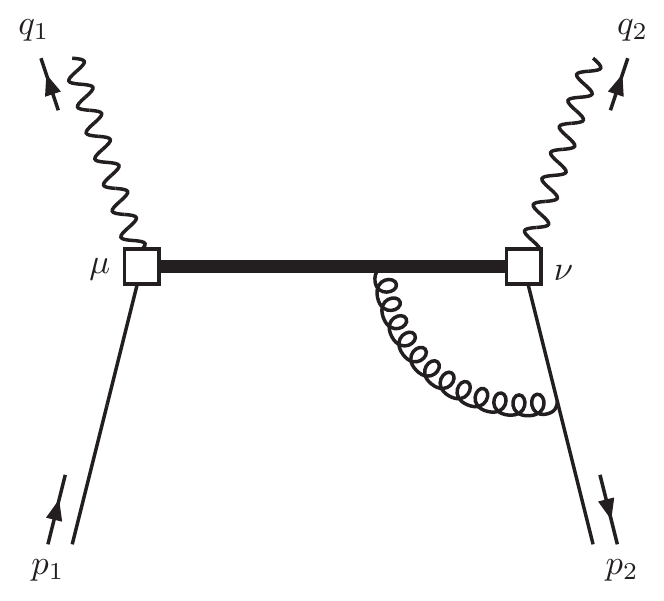}
}
\subfigure[]{\label{fig:self2}
	\includegraphics[width=0.25\linewidth]{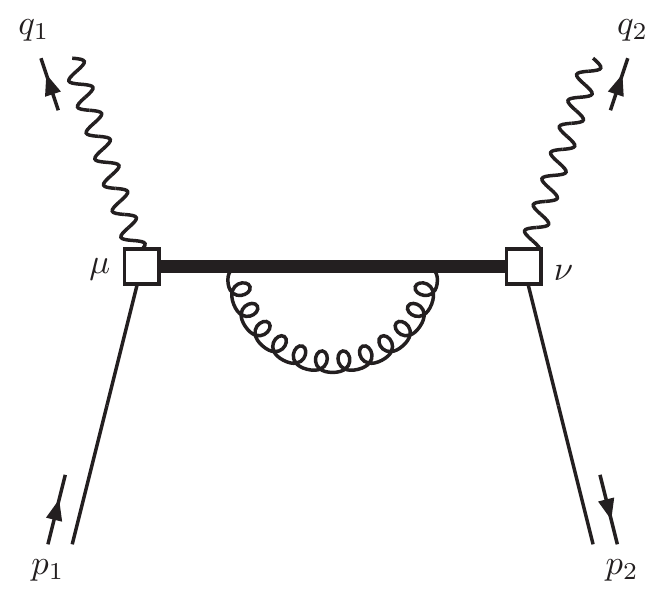}
}
	\caption{The hard scattering kernel up to $\mathcal{O}(\alpha_s)$. One must also sum crossed versions of these diagrams.}
	\label{fig:oneloop}
\end{figure}

We carry out calculations of such Wilson coefficients with asymptotic massless external quarks, where both the ultraviolet (UV) and infrared (IR, or collinear) divergences are regulated using dimensional regularization. The OPEs for the PDF and LCDA cases expand in terms of Mellin and Gegenbauer moments, respectively, so we treat them differently with different external states. The Feynman diagrams contributing to the one-loop corrections are shown in \figs{oneloop}{self}.

\begin{figure}
	\centering
	\includegraphics[width=0.3\linewidth]{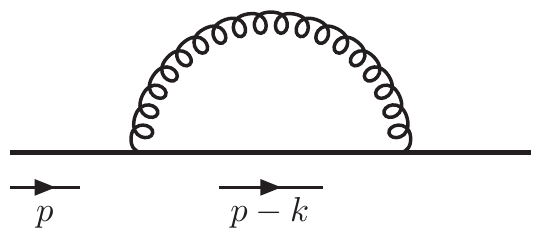}
	\caption{Self energy correction to the light quark.}
	\label{fig:self}
\end{figure}

\subsection{Parton Distribution Functions}
\label{sec:nlopdf}
 
 The physical process that corresponds to the hadronic amplitude for the PDF case is heavy quark production through the charged current in neutrino DIS~\cite{Gottschalk:1980rv,Gluck:1996ve}, except that the latter only involves left-handed fermions. The next-to-leading order correction to the cross section was first calculated in Ref.~\cite{Gottschalk:1980rv}, and a mistake in that calculation was corrected in Ref.~\cite{Gluck:1996ve}. Currently, the QCD correction has been computed up to ${\cal O}(\alpha_s^3)$~\cite{Blumlein:2011zu,Hasselhuhn:2013swa,Blumlein:2014fqa,Behring:2015roa,Behring:2016hpa}, and power corrections due to heavy quark mass have also been obtained at ${\cal O}(\alpha_s^2)$~\cite{Blumlein:2016xcy}.
 In principle we can derive the dispersion relation~\cite{Christ:1972ms} for HOPE and the neutrino DIS structure functions, which can be used to extract the Wilson coefficients~\cite{Blumlein:2011zu}.
Nevertheless, in this work we directly calculate the Wilson coefficients from the hadronic amplitude using the standard method in field theory. To be specific, we first compute the one-loop quark matrix element of the current-current correlator, and then compare the coefficients in its Taylor expansion in $\tilde \omega$ to the matrix elements of the corresponding Mellin moments. The IR divergences between the two at each order of $\tilde \omega$ must be identical, and their difference gives the perturbative Wilson coefficient. In the end, we compare our results to those extracted from the one-loop neutrino DIS cross section formula~\cite{Gottschalk:1980rv,Gluck:1996ve,Blumlein:2011zu}.

For the PDF case, we set $p_1=p_2=p$ in \fig{oneloop}. The tree-level 
hadronic amplitude is 
\begin{align}
T_q^{\mu\nu}(q,p) &= \int d^4x\ e^{iq\cdot x}\langle p| {\cal T}\left\{J^\mu_A({x/2})J^\nu_A(-{x/2})\right\}|p\rangle\nn\\
	&=i \bar{u}(p) \left[\gamma^\mu{ \slashed q +  \slashed p+m_\Psi\over (q + p)^2 \!- \! m_\Psi^2}\gamma^\nu  \!+\! \gamma^\nu {-\slashed q + \slashed p+m_\Psi\over (q - p)^2 \!-\! m_\Psi^2}\gamma^\mu \right]u(p)\,,
\end{align}
where $|p\rangle$ is a free quark state and $u(p)$ is a quark Dirac spinor.
Due to chiral symmetry, the terms linear in $m_\Psi$ vanish between the light-quark spinors as they are proportional to $\bar{u}(p)\gamma^{\mu}\gamma^\nu u(p)$ or $\bar{u}(p)\gamma^{\nu}\gamma^\mu u(p)$, and therefore 
\begin{align}\label{eq:unpol}
	T_q^{\{\mu\nu\}}(q,p) &= {T_q^{\mu\nu}(q,p)+T_q^{\nu\mu}(q,p)\over2}\nn\\
	&=  2i \bar{u}(p)q^{\{\mu} \gamma^{\nu\}}u(p) \left[{ 1 \over (q + p)^2 \!- \! m_\Psi^2} \!-\! {1 \over (q - p)^2 \!-\! m_\Psi^2} \right] + 2i \bar{u}(p)p^{\{\mu} \gamma^{\nu\}} u(p) \left[{ 1 \over (q + p)^2 \!- \! m_\Psi^2} \!+\! {1 \over (q - p)^2 \!-\! m_\Psi^2} \right]\nn\\
	&\quad  -ig^{\mu\nu} \bar{u}(p) \slashed q u(p) \left[{ 1 \over q^2+2q\cdot p \!- \! m_\Psi^2} \!-\! {1 \over q^2-2q\cdot p \!-\! m_\Psi^2} \right]\nn\\
	&= -4i\sum_{\substack{n=1,\\\rm odd}}^\infty \left({2 q_{\mu_1}\over \tilde Q^2}\right)\cdots \left({2 q_{\mu_n}\over \tilde Q^2}\right) \langle p| q^{\{\mu}{\cal O}_{n+1,V}^{\nu \mu_1\ldots \mu_n\}}|p\rangle - 4i\sum_{\substack{n=0,\\\rm even}}^\infty \left({2 q_{\mu_1}\over \tilde Q^2}\right)\cdots \left({2 q_{\mu_n}\over \tilde Q^2}\right) \langle p| {\cal O}_{n+2,V}^{\mu\nu \mu_1\ldots \mu_n}|p\rangle \nn\\
	&\quad + ig^{\mu\nu}\sum_{\substack{n=2,\\\rm even}}^\infty \left({2 q_{\mu_1}\over \tilde Q^2}\right)\cdots \left({2 q_{\mu_n}\over \tilde Q^2}\right) \langle p| {\cal O}_{n,V}^{\mu_1\ldots \mu_n}|p\rangle \,.
\end{align}

Since the heavy-to-light currents we use are not conserved due to the nonzero quark mass, the hadronic amplitude $T_q^{\{\mu\nu\}}(q,p)$ does not satisfy the chiral Ward identity,
\begin{align}
	q_\mu T_q^{\{\mu\nu\}}(q,p) \neq 0\,.
\end{align}
Therefore,  $T_q^{\{\mu\nu\}}(q,p)$ can include up to four independent Lorentz tensor structures,
\begin{align}
	g^{\mu\nu}\,,\ \ p^\mu p^\nu\,,\ \ (p^\mu q^\nu + p^\nu q^\mu)\,,\ \ q^\mu q^\nu\,,
\end{align}
where the first three structures are present at tree level, as seen in \eq{unpol}, while the last one enters at ${\cal O}(\alpha_s)$.

In perturbative QCD, the hadronic amplitude $T_q^{\{\mu\nu\}}$ includes collinear divergences which are regulated by dimensional regularization with $d=4-2\epsilon$. 
At one-loop order, all UV divergences in the Feynman diagrams in \figs{oneloop}{self} cancel, except for that from the heavy quark mass correction. Therefore, the hadronic amplitude is independent of the wave function renormalization scheme for the quarks, but depends on the scheme for heavy-quark mass renormalization. In our calculation, we consider the heavy-quark mass in both on-shell and $\MS$ schemes.

The OPE of the hadronic amplitude $T_q^{\{\mu\nu\}}$ to all orders in perturbation theory takes the form~\cite{Gottschalk:1980rv}
\begin{align}\label{eq:tnlo}
T_q^{\{\mu\nu\}}(q,p,m_\Psi,\mu_{\tiny\rm uv},\mu_{\tiny\rm ir},\epsilon_{\tiny\rm ir})
&={4i \bar{u}(p)q^{\{\mu} \gamma^{\nu\}} u(p) \over \tilde Q^2}\sum_{\substack{n=1,\\\rm odd}}^\infty \tilde \omega^{n}C_{1,n}(\tilde Q^2, \tau,\mu^2_{\rm uv}) a_{n,V}(\eps_{\tiny \rm ir},\mu^2_{\rm uv},\mu^2_{\rm ir}) \nn\\
&\quad+{4 i \bar{u}(p)p^{\{\mu} \gamma^{\nu\}}u(p) \over \tilde Q^2} \sum_{\substack{n=1,\\\rm odd}}^\infty \tilde \omega^{n-1}C_{2,n}(\tilde Q^2, \tau,\mu^2_{\rm uv}) a_{n,V}(\eps_{\tiny \rm ir},\mu^2_{\rm uv},\mu^2_{\rm ir}) \nn\\
&\quad- {2ig^{\mu\nu}\bar{u}(p) \slashed q u(p)\over \tilde Q^2} \sum_{\substack{n=1,\\\rm odd}}^\infty \tilde \omega^{n}C_{3,n}(\tilde Q^2, \tau,\mu^2_{\rm uv}) a_{n,V}(\eps_{\tiny \rm ir},\mu^2_{\rm uv},\mu^2_{\rm ir})\nn\\
&\quad + {2i q^\mu q^\nu  \bar{u}(p) \slashed q u(p) \over \tilde Q^4} \sum_{\substack{n=1,\\\rm odd}}^\infty \tilde \omega^{n}C_{4,n}(\tilde Q^2, \tau,\mu^2_{\rm uv}) a_{n,V}(\eps_{\tiny \rm ir},\mu^2_{\rm uv},\mu^2_{\rm ir}) \,,
\end{align}
where $C_{i,n}(\tilde Q^2, \tau,\mu^2_{\tiny\rm uv})$ are the Wilson coefficients, and $a_{n,V}(\eps_{\tiny \rm ir},\mu^2_{\rm uv},\mu^2_{\rm ir})$ are the Mellin moments in perturbation theory. We use $\eps_{\tiny \rm uv}$ and $\eps_{\tiny \rm ir}$ to differentiate the UV and IR divergences, and $\mu_{\tiny\rm uv}$ and $\mu_{\tiny\rm ir}$ are the $\MS$ scales for UV renormalization and IR regularization.
The Mellin moments in the massless quark state are
\begin{align}\label{eq:moment1}
	a_{n,V}(\eps_{\rm ir},\mu^2_{\rm uv},\mu^2_{\rm ir}) &=1+  {\alpha_s C_F\over 4\pi}\gamma_n^{(0)}\Big({1\over\epsilon'_{\tiny\rm ir}} - \ln{\mu^2_{\tiny\rm uv} \over \mu^2_{\tiny\rm ir}}\Big) + {\cal O}(\alpha_s^2)\,,
\end{align}
where $1/\eps'_{\tiny\rm ir}=1/\eps_{\tiny\rm ir}-\gamma_E+ \ln(4\pi )$, and the factor
\begin{align}
	\gamma_n^{(0)} &= 2\left[{3+2n\over 2+3n+n^2} + 2H_n -{3\over2} \right]
\end{align}
is the minus one-loop anomalous dimension of $a_{n,V}$
with the harmonic number
\begin{align}
	H_n=\sum_{j=1}^n {1\over j}\,.
\end{align}
Note that the regulator $\eps_{\tiny \rm ir}$ is introduced to bookkeep the IR divergence in our perturbative calculation, but it is not a prediction for the IR contribution in the hadron state, which is intrinsically nonperturbative. Due to the cancellation of $\eps_{\tiny \rm ir}$-dependence between the hadronic amplitude and the Mellin moments, the Wilson coefficients only depend on the UV scales and are thus perturbatively calculable, which allows us to make predictions for the Mellin moments once the hadronic amplitude are computed from lattice QCD.

For convenience, from now on we will not differentiate $\eps_{\tiny \rm uv}$ and $\eps_{\tiny \rm ir}$, and we set $\mu_{\tiny \rm uv}=\mu_{\tiny \rm ir}=\mu$, which is the factorization scale. Therefore, the hadronic amplitude is denoted as $T_q^{\{\mu\nu\}}(q,p,m_\Psi,\mu,\epsilon)$, while the Mellin monents are denoted $a_{n,V}(\eps)$.\\

The Wilson coefficients $C_{i,n}(\tilde Q^2, \tau,\mu^2)$ are also series expansions in $\alpha_s$ given by
\begin{align}\label{eq:cnexp1}
	C_{i,n}& = 1 + {\alpha_sC_F\over 4\pi}C_{i,n}^{(1)}+{\cal O}(\alpha_s^2)\,,\ \ \ {\rm for }\ i=1\,,2\,,3\,,\\
	C_{4,n}& = {\alpha_sC_F\over 4\pi}C_{4,n}^{(1)}+{\cal O}(\alpha_s^2)\,.
\end{align}
Therefore, at ${\cal O}(\alpha_s)$,
\begin{align}\label{eq:nlo2}
C_{i,n} a_{n,V}(\eps) &= 1 +  {\alpha_s C_F\over 4\pi}\left[ {\gamma_{n}^{(0)}\over\epsilon'} + C^{(1)}_{i,n}\right]+ {\cal O}(\alpha_s^2) \,,
\end{align}
for $i=1,2,3$, and
\begin{align}\label{eq:nlo3}
	C_{4,n} a_{n,V}(\eps) &= {\alpha_s C_F\over 4\pi} C^{(1)}_{4,n} \!+\! {\cal O}(\alpha_s^2) \,.
\end{align}

The one-loop Wilson coefficients $C^{(1)}_{i,n}(\tilde Q^2, \tau,\mu^2)$ have the structure
\begin{align}\label{eq:cnexp2}
	C^{(1)}_{i,n}(\tilde Q^2, \tau,\mu^2) &= \gamma_{n}^{(0)} \ln{\mu^2\over \tilde Q^2} \!+\! R_{i,n}^{(1)}(\tilde Q^2, \mu^2, \tau)\,,\ {\rm for}\ i=1\,,2\,,3\,,\\
	C^{(1)}_{4,n}(\tilde Q^2, \tau,\mu^2) &=  R_{4,n}^{(1)}(\tau)\,,
\end{align}
following the renormalization group equation of the Mellin moments. Therefore, by Taylor expanding the one-loop matrix element of the hadronic amplitude $T_q^{\{\mu\nu\}}(q,p,m_\Psi,\mu,\epsilon)$ in $\tilde \omega$, we can read off the finite term $R_{i,n}^{(1)}(\tau)$ by comparing to \eqs{tnlo}{nlo2}. This can be achieved by employing the formulae provided in the supplementary material.

The full one-loop result of $T_q^{\{\mu\nu\}}(q,p,m_\Psi,\mu,\epsilon)$ can be found in \app{pdf}. Here, we provide the Wilson coefficients for the first few moments.

In the on-shell scheme for heavy-quark mass renormalization, which we denote by ``OS'' in the superscripts of $R^{{\rm OS}}_{i,n}(\tau)$ below, there is no dependence on $\ln(\mu^2/\tilde Q^2)$, and
\begin{align}
	R_{1,1}^{{\rm OS}(1)}(\tau) &= - {12-22 \tau+20 \tau^2-11 \tau^3 \over 3\tau^3}  - {2(1-\tau)(6 - 8 \tau + 7 \tau^2 + 13 \tau^3)\ln(1-\tau) \over 3\tau^4}\,,\\
	R_{1,3}^{{\rm OS}(1)}(\tau) &= -{480 - 696 \tau + 262 \tau^2 + 173 \tau^3 + 332 \tau^4 - 
		824 \tau^5 \over 60\tau^5}  \nn\\
	&\qquad -  {(1-\tau)(240 - 228 \tau + 57 \tau^2 + 97 \tau^3 + 217 \tau^4 + 
		337 \tau^5) \ln(1-\tau) \over 30 \tau^6}\,.
\end{align}

\begin{align}
	R_{2,1}^{{\rm OS}(1)}(\tau) &= - {2 - 5 \tau + 2 \tau^2 \over 3\tau^2 } - {2(1-\tau)(1 - 2 \tau + 10 \tau^2) \ln(1-\tau) \over 3\tau^3} \,,\\
	R_{2,3}^{{\rm OS}(1)}(\tau) &=- {144 - 198 \tau + 233 \tau^2 + 226 \tau^3 - 678 \tau^4 \over 60\tau^4} - {(1-\tau)(72 - 63 \tau + 97 \tau^2 + 157 \tau^3 + 277 \tau^4) \ln(1-\tau) \over 30 \tau^5}\,.
\end{align}

\begin{align}
	R_{3,1}^{{\rm OS}(1)}(\tau) &=  {6 - 11 \tau - 5 \tau^2 + 7 \tau^3 \over 3\tau^3} + {2(1-\tau)(3 - 4 \tau - 4 \tau^2 - 13 \tau^3)\ln(1-\tau) \over 3\tau^4}\,,\\
	R_{3,3}^{{\rm OS}(1)}(\tau) &= {120 - 324 \tau - 2 \tau^2 - 103 \tau^3 - 182 \tau^4 + 
		716 \tau^5 \over 60\tau^5}  \nn\\
	&\qquad +   {(1-\tau)(60 - 132 \tau - 57 \tau^2 - 97 \tau^3 - 157 \tau^4 - 
		337 \tau^5) \ln(1-\tau) \over 30 \tau^6}\,.
\end{align}

\begin{align}
	R_{4,1}^{{\rm OS}(1)}(\tau) &= - {4(12 - 18 \tau + 4 \tau^2 + \tau^3) \over 3\tau^4} - {16(1-\tau)^2\ln(1-\tau) \over \tau^5}\,,\\
	R_{4,3}^{{\rm OS}(1)}(\tau) &= - {2(60 - 90 \tau + 20 \tau^2 + 5 \tau^3 + 2 \tau^4 + \tau^5) \over 5\tau^6}  - {24(1-\tau)^2\ln(1-\tau) \over \tau^7}\,.
\end{align}

The $\MS$ Wilson coefficients are related to the on-shell scheme ones through
\begin{align}
	R_{1,n}^{\MS (1)}(\tilde Q^2, \mu^2, \tau) - R_{1,n}^{{\rm OS}(1)}(\tau) &= R_{3,n}^{\MS (1)}(\tilde Q^2, \mu^2, \tau) - R_{3,n}^{{\rm OS}(1)}(\tau) = -2(n+1)(1-\tau)\Big(4 + 3\ln{\mu^2\over m_\Psi^2}\Big)\,,\\
	R_{2,n}^{\MS (1)}(\tilde Q^2, \mu^2, \tau) - R_{2,n}^{{\rm OS}(1)}(\tau) &= -2n(1-\tau)\Big(4 + 3\ln{\mu^2\over m_\Psi^2}\Big)\,, \\
	R_{4,n}^{\MS (1)}(\tau) - R_{4,n}^{{\rm OS}(1)}(\tau) &= 0\,.
\end{align}

Finally, we compare our results to those derived from the Mellin moments of the hard coefficient functions in Eq.~(42) of Ref.~\cite{Gottschalk:1980rv} for heavy-quark production from neutrino DIS,  with a correction in Ref.~\cite{Gluck:1996ve}. Both $R_{2,n}^{{\rm OS}(1)}(\tau)$ and $R_{3,n}^{{\rm OS}(1)}(\tau)$ were presented as $h_2^q(\tau,n+2)$ and $h_1^q(\tau,n+2)$ in compact forms for arbitrary $n$ in Ref.~\cite{Blumlein:2011zu}. We found agreement in the on-shell scheme for all the $R_{i,n}^{{\rm OS}(1)}(\tau)$, except that the factor $A_5$ in Table I in Ref.~\cite{Gottschalk:1980rv} must be corrected by a factor of two. Since the factor $A_2$ in Table I in Ref.~\cite{Gottschalk:1980rv} is also missing a factor of two, as was pointed out in Ref.~\cite{Gluck:1996ve}, we believe that $A_5$ has the same problem.

In the heavy-quark massless limit $m_\Psi\to0$ or $\tau\to1$, the above results no longer depend on the heavy-quark mass renormalization scheme and agree with those in the literature~\cite{Gross:1973ju,Gross:1974cs,Georgi:1951sr,Calvo:1977ba,DeRujula:1976baf,Altarelli:1978id,Bardeen:1978yd},
\begin{align}
	R_{1,n}^{(1)}(1) &= R_{2,n}^{(1)}(1) = 3H_{n+1} - 4 H_{n+1}^{(2)} - {2H_{n+1}\over (n+1)(n+2)} + 4 \sum_{s=1}^{n+1}{1\over s} \sum_{j=1}^s {1\over j} + {3\over n+1} + {4\over n+2} + {2\over (n+1)^2 } -9 \,,\nn\\
	R_{3,n}^{(1)}(1) &= R_{1,n}^{(1)}(1)  - {4\over n+2}\,,\ \ \ \ \ \ \ \ \ R_{4,n}^{(1)}(\tau) = {4\over n+2}\,,
\end{align}
where
\begin{align}
	H_n^{(2)}=\sum_{j=1}^n {1\over j^2}\,.
\end{align}

\subsection{Helicity Parton Distribution Functions}
\label{sec:helicity}

To calculate the helicity PDF, we use the anti-symmetrized hadronic amplitude
\begin{align}
	T_q^{[\mu\nu]}(q,p) &= {T_q^{\mu\nu}(q,p)-T_q^{\nu\mu}(q,p)\over2}\,.
\end{align}
The HOPE of $T^{[\mu\nu]}_q$ takes a similar form to \eq{tnlo},
\begin{align}\label{eq:tnlo2}
T_q^{[\mu\nu]}(q,p,m_\Psi,\mu,\epsilon)&= -{2\epsilon^{\mu\nu\rho\sigma}q_\rho  \over \tilde Q^2} \ \bar{u}(p)\gamma_\sigma \gamma_5 u(p) \sum_{\substack{n=0,\\\rm even}}^\infty  \tilde \omega^n  \tilde C_n(\tilde Q^2, \tau,\mu^2) a_{n,A}(\epsilon)\,.
\end{align}
The Wilson coefficients $\tilde C_n$ have the same perturbative expansion as in \eqs{cnexp1}{cnexp2}, i.e.,
\begin{align}
	\tilde C^{(1)}_{n}(\tilde Q^2, \tau,\mu^2) &= \gamma_{n}^{(0)} \ln{\mu^2\over \tilde Q^2} + \tilde R^{(1)}_{n}(\tilde Q^2, \mu^2, \tau)\,,
\end{align}
so they can be extracted following exactly the same procedure as the unpolarized PDF.

The full one-loop result of $T_q^{[\mu\nu]}(q,p,m_\Psi,\mu,\epsilon)$ can be found in \app{hpdf}. Here we provide results for the lowest four Wilson coefficients:
\begin{align}
	\tilde R^{{\rm OS}(1)}_0(\tau) &=-3 -\frac{6 (1-\tau ) \ln (1-\tau )}{\tau}\,,\\
	\tilde R^{{\rm OS}(1)}_2(\tau) &= -\frac{18 + 17 \tau + 34 \tau^2 - 84 \tau^3}{12 \tau ^3} -\frac{(1-\tau)(9 + 13 \tau + 25 \tau^2 + 61 \tau^3) \ln (1-\tau )}{6 \tau ^4}\,,\\
	\tilde R^{{\rm OS}(1)}_4(\tau) &= -\frac{300 + 6 \tau + 118 \tau^2 + 192 \tau^3 + 294 \tau^4 - 
		1433 \tau^5}{90 \tau ^5} \nn\\
	&\qquad - \frac{(1-\tau)(50 + 26 \tau + 41 \tau^2 + 61 \tau^3 + 91 \tau^4 + 
		181 \tau^5) \ln (1-\tau )}{15 \tau ^6}\,,
\end{align}
and 
\begin{align}
	\tilde R^{\MS (1)}_n(\tilde Q^2, \mu^2, \tau) - \tilde R^{{\rm OS}(1)}_n(\tau) &= -2(n+1)(1-\tau)\Big(4 + 3\ln{\mu^2\over m_\Psi^2}\Big) \,.
\end{align}

Again, we also find agreement between our results and those from the neutrino DIS cross section for heavy-quark production~\cite{Gottschalk:1980rv,Gluck:1996ve,Blumlein:2011zu} in the on-shell scheme. In particular, the reader can find a compact form of $\tilde R_{n}^{{\rm OS}(1)}(\tau)$ for arbitrary $n$ in Ref.~\cite{Blumlein:2011zu}, where it was presented as $h_3^q(\tau,n+1)$.

In the massless limit $m_\Psi\to0$ or $\tau\to1$, the above results become independent of the heavy-quark mass renormalizaiton scheme and agree with those in the literature~\cite{Kodaira:1978sh,Kodaira:1979ib,Kodaira:1979pa},
\begin{align}\label{eq:forwardpdf}
	\tilde R^{(1)}_n(1) & = 2H_{n+1}^2 - 2H_{n+1}^{(2)} + 3H_{n+2} + {3-2H_n\over (n+1)(n+2)} - 9\,.
\end{align}

\subsection{Light-Cone Distribution Amplitude}
\label{sec:danlo}

For the pseudoscalar meson LCDA defined in \eq{lcda}, we calculate the transition amplitude of the heavy-quark hadronic amplitude from vacuum to the asymptotic state composed of a quark and antiquark, that is,
\begin{equation}
	\ket{\pi(\mathbf{p})}\rightarrow |u({1\over2}(1+x_0)p,\uparrow) \bar{d}({1\over2}(1-x_0)p,\downarrow)\rangle
\end{equation}
with $-1< x_0 < 1$.

In \fig{oneloop}, this corresponds to choosing
\begin{equation}
p_1={1\over2}(1+x_0)p \,, \ \ \ \ \ \ \ \ \ p_2=-{1\over2}(1-x_0)p\,,
\end{equation}
so the external line carrying $p_2$ stands for the antiquark. In this simple model, the LCDA at tree level is
\begin{align}
	\phi_\pi^{(0)}(x,x_0) &= \delta(x-x_0)\,,
\end{align}

The Gegenbauer moments~\cite{Radyushkin:1977gp} for fixed $x_0$ are therefore
\begin{align}
	\phi_n^{(0)}(x_0) &\equiv  {4(n+{3\over2})\over 3(n+1)(n+2)}\int_{-1}^{1} dx\ {\cal C}_n^{3/2}(x) \phi_\pi^{(0)}(x,x_0) = {4(n+{3\over2})\over 3(n+1)(n+2)} {\cal C}_n^{3/2}(x_0) \,.
\end{align}

At tree level, the hadronic amplitude
\begin{align}
	V_q^{\mu\nu(0)}(q,p)
	&=\bar{v}({1\over2}(1-x_0)p,\downarrow)\left[\gamma^\mu\gamma_5 {i\over \slashed q + {x_0\slashed p\over2}-m_\Psi}\gamma^\nu\gamma_5  + \gamma^\nu\gamma_5 {i\over  -\slashed q +{x_0\slashed p\over2}-m_\Psi}\gamma^\mu\gamma_5 \right]u({1\over2}(1+x_0)p,\uparrow)\,.
\end{align}

After antisymmetrization of $\mu$ and $\nu$, the conformal or Gegenbauer OPE of the hadronic amplitude is~\cite{Braun:2007wv},
\begin{align}
	V_q^{[\mu\nu](0)}(q,p,m_\Psi,x_0)
	&=  \epsilon^{\mu\nu\rho\sigma}q_\rho \bar{v}\gamma_\sigma \gamma_5 u \left[{1\over q^2 + x_0 p\cdot q-m_\Psi^2} +{1\over q^2 -x_0p\cdot q-m_\Psi^2}\right]\nn\\
	&=-{2\epsilon^{\mu\nu\rho\sigma}q_\rho \over \tilde{Q}^2}\bar{v}\gamma_\sigma \gamma_5 u \sum_{\substack{n=0,\\\rm even}}^\infty  {\cal F}^{(0)}_n(\tilde \omega) \phi_n^{(0)}(x_0)\,,
\end{align}
where
\begin{align}
	{\cal F}^{(0)}_n(\tilde \omega) &={3\over2}\frac{\sqrt{\pi } (n+1) (n+2) n! }{2^{n+2} \Gamma \left(n+\frac{5}{2}\right)}\left(\frac{\tilde \omega}{2}\right)^n  \, _2F_1\left(\frac{n+1}{2},\frac{n+2}{2};n+\frac{5}{2};\frac{\tilde \omega^2}{4}\right)\,.
\end{align}

At one-loop order, the Gegenbauer moments and the Mellin moments, as defined in \eq{phiM}, share the same anomalous dimensions.
Therefore,
	\begin{align}
		\phi_n (x_0,\eps) &= \phi_n^{(0)}(x_0) \left[1+{\alpha_s C_F\over 4\pi} {\gamma_n^{(0)}\over\epsilon'} + {\cal O}(\alpha_s^2)\right]\,.
\end{align}

Using the renormalization group equation of the Gegenbauer moments, the hadronic amplitude should be expanded in the same fashion as that in Eq.~\eqref{eq:tnlo},
\begin{align}
	&V_q^{[\mu\nu]}(q,p,m_\Psi,x_0,\mu,\eps)  \nn\\
	&=-{2 \epsilon^{\mu\nu\rho\sigma}q_\rho \over \tilde{Q}^2}\bar{v}\gamma_\sigma \gamma_5 u \sum_{\substack{n=0,\\\rm even}}^\infty {\cal F}_n(\tilde Q^2,\mu^2,\tilde \omega,\tau) \phi_n(x_0,\eps)\nn\\
	&=-{2 \epsilon^{\mu\nu\rho\sigma}q_\rho \over \tilde{Q}^2}\bar{v}\gamma_\sigma \gamma_5 u \sum_{\substack{n=0,\\\rm even}}^\infty \Bigg[ {\cal F}^{(0)}_n(\tilde \omega) \Big(1 + {\alpha_sC_F\over 4\pi}\gamma_n^{(0)}\Big({1\over \eps'} + \ln{\mu^2\over \tilde Q^2} \Big)\Big) +\frac{\alpha_S C_F}{4\pi}{\cal R}^{(1)}_n(\tilde Q^2, \mu^2, \tau,\tilde \omega)\Bigg] \phi_n^{(0)}(x_0) +{\cal O}(\alpha_s^2)\,,
\end{align}
where the coefficient function ${\cal F}_n$ is expanded to one-loop order as
\begin{align}
		{\cal F}_n(\tilde Q^2,\mu^2,\tilde \omega,\tau) &= {\cal F}^{(0)}_n(\tilde \omega) \left[1+ {\alpha_s C_F\over 4\pi} \gamma_n^{(0)} \ln{\mu^2\over \tilde Q^2} \right]+{\alpha_s C_F\over 4\pi} {\cal R}^{(1)}_n(\tilde Q^2, \mu^2, \tau,\tilde \omega)  +{\cal O}(\alpha_s^2)\,.
\end{align}

As one can see, the logarithmic part of ${\cal F}_n$ is proportional to the tree-level coefficient function ${\cal F}^{(0)}_n(\tilde \omega) $, because at leading logarithmic accuracy QCD is conformal and the Gegenbauer moments do not mix with each other. 
Beyond leading logarithm, the conformal symmetry is broken, so the Gegenbauer moments start to mix. 

In the conformal approximation, the OPE of the current-current correlator for the pion LCDA has been derived for massless quarks in Refs.~\cite{Mueller:1997ak,Braun:2003rp,Braun:2007wv}. The coefficient functions in the conformal OPE are
\begin{align}
	{\cal F}^{\rm conf}_n(Q^2,\mu^2,\tilde \omega,\delta) &= {3\over2}\frac{\sqrt{\pi } (n+1) (n+2) n! }{2^{n+2} \Gamma \left(n+\frac{5}{2}\right)}\left(\frac{\tilde \omega}{2}\right)^n \left({\mu^2\over Q^2}\right)^{\delta} \ _2F_1\left({n+1+\delta\over2}, {n+2+\delta\over2}, n+{5\over2}+\delta, {\omega^2\over4}\right)c_n(\alpha_s)\,,
\end{align}
where the logarithms of $\mu^2/Q^2$ have been resummed with the anomalous dimension
\begin{align}
	\delta = {\alpha_s\over 4\pi}\gamma^{(0)}_n+ {\cal O}(\alpha_s^2)\,,
\end{align}
and the factor
\begin{align}
	c_n(\alpha_s) &= 1 + {\alpha_sC_F\over 4\pi}\tilde R^{(1)}_n(1) + {\cal O}(\alpha_s^2)\,,
\end{align}
with $\tilde R^{(1)}_n(1)$ beging the same as those in the Wilson coefficients for the helicity PDF case~\cite{Kodaira:1978sh,Kodaira:1979ib,Kodaira:1979pa}, which has been given in \eq{forwardpdf}.

If we expand ${\cal F}^{\rm conf}_n(Q^2,\mu^2,\tilde \omega,\delta)$ to ${\cal O}(\alpha_s)$,
\begin{align}
	{\cal F}^{\rm conf}_n(Q^2,\mu^2,\tilde \omega,\delta) &= {\cal F}^{\rm conf}_n(Q^2,\mu^2,\tilde \omega,0) + \delta {\partial \over \partial \delta'}{\cal F}^{\rm conf}_n(Q^2,\mu^2,\tilde \omega,\delta')\Big|_{\delta'=0} + {\cal O}(\alpha_s^2)\,,
\end{align}
and compare it to ${\cal F}_n(\tilde Q^2,\mu^2,\tilde \omega,\tau)$ in the massless heavy-quark limit $\tau\to1$, we find that they are not equal, which leads to the breaking of conformal symmetry in addition to the nonzero $\beta$-function that enters at ${\cal O}(\alpha_s^2)$.
Such conformal symmetry breaking effects in the $\MS$ scheme have been observed in the literature~\cite{Melic:2002ij}, and it was proposed to rotate the $\MS$ factorization scheme to the so-called ``conformal subtraction'' scheme where the new Gegenbauer moments evolve autonomously without mixing~\cite{Melic:2002ij}.
Nevertheless, the numerical difference between ${\cal F}_n(\tilde Q^2,\mu^2,\tilde \omega,1) $ and ${\cal F}^{\rm conf}_n(Q^2,\mu^2,\tilde \omega,\delta) $ turns out to be small. Since in our calculation we use a massive quark, the conformal symmetry is explicitly broken anyway, so we use the exact $\MS$ result of ${\cal F}_n(\tilde Q^2,\mu^2,\tilde \omega,\tau)$.\\

The full one-loop correction for the hadronic amplitude from \figs{oneloop}{self} is
\begin{align}\label{eq:oneloop}
V_q^{[\mu\nu](1)}(q,p,m_\Psi,x_0,\mu,\eps)
	&= - {\alpha_sC_F\over 4\pi} {2 \epsilon^{\mu\nu\rho\sigma}q_\rho \over \tilde{Q}^2}\bar{v}\gamma_\sigma \gamma_5 u  \Big[\Big({1\over \epsilon'} + \ln{\mu^2\over \tilde Q^2}\Big) F^{(1)}(\tilde \omega, x_0) + {\mathbf R}^{(1)}(\tilde Q^2, \mu^2, \tau,\tilde \omega,x_0)\Big]\,,
\end{align}
where
\begin{align}\label{eq:lcdall}
	F^{(1)}(\tilde \omega, x_0)  &={4\over (1 - x_0^2) \tilde \omega^2 (4 - x^2_0 \tilde \omega^2)}\Bigg\{3 \left(x_0^2-1\right) \tilde \omega^2 +\big[\left(x^2_0-2\right) \tilde \omega^2+4\big] \ln \frac{ 4-x_0^2 \tilde \omega^2}{4-\tilde \omega^2}\nn\\
	&\qquad +\tilde \omega \Big[\left(4- x_0^2 \tilde \omega^2\right)\tanh ^{-1}\big(\frac{\tilde \omega}{2}\big)+ x_0\left((2- x_0^2) \tilde \omega^2-4\right) \tanh ^{-1}\big({x_0 \tilde \omega\over2}\big)\Big]\Bigg\}\,,
\end{align}
and the finite part ${\mathbf R}^{(1)}$ is provided in the supplementary material. The details of calculating $V_q^{[\mu\nu](1)}(q,p,m_\Psi,x_0,\mu,\eps)$ in both the on-shell and $\MS$ schemes can be found in \app{lcda}.

Using the orthonomality of Gegenbauer polynomials, we can verify that
\begin{align}
	{\cal F}^{(0)}_n(\tilde \omega)\gamma_n^{(0)} & ={3\over4} \int_{-1}^{1}dx_0 (1-x_0^2) F^{(1)}(\tilde \omega, x_0) {\cal C}_n^{3/2}(x_0)\,,
\end{align}
and obtain ${\cal R}^{(1)}_n(\tilde Q^2, \mu^2, \tau,\tilde \omega)$ as
\begin{align}\label{eq:Rn1}
	{\cal R}^{(1)}_n(\tilde Q^2, \mu^2, \tau, \tilde \omega) & ={3\over4} \int_{-1}^{1}dx_0\ (1-x_0^2) {\mathbf R}^{(1)}(\tilde Q^2, \mu^2, \tau, \tilde \omega, x_0) {\cal C}_n^{3/2}(x_0)\,,
\end{align}

The results of ${\mathbf R}^{(1)}(\tilde Q^2, \mu^2, \tau, \tilde \omega, x_0)$ and ${\cal R}^{(1)}_n(\tilde Q^2, \mu^2, \tau,\tilde \omega)$ are new in this work.

\section{Conclusion}
\label{sec:summary}

This paper contains novel and important information for applying the HOPE strategy to determine directly the $\xi$-dependence and moments for the PDFs and LCDAs. In particular, we discuss the theoretical details of the HOPE method applied to the PDF and LCDA.  After introducing the unpolarised PDF, helicity PDF and LCDA we begin with a general overview of the HOPE strategy. This approach is a modified OPE relation where the conventional current operators are replaced with heavy-light flavour changing currents.  Importantly the use of a fictitious heavy quark species allows for suppression of higher-twist effects. In the regime $\tilde{Q}\sim m_\Psi$, the effects of the heavy quark mass can be incorporated into the Wilson coefficients, and thus the resulting moments are the standard QCD moments.  
 The HOPE method was first proposed in Ref.~\cite{Detmold:2005gg} for computing higher moments of the PDFs with lattice QCD.  The results presented in this article extend this proposal for direct determinations of $\xi$-dependence of the PDFs and LCDAs.  As demonstrated in Sec.~\ref{sec:Hope_qPDF}, the PDFs and LCDAs can be related to appropriate HOPE hadronic amplitudes through a perturbative matching condition.   This offers a novel, future direction for analysing these HOPE hadronic amplitudes.

Our work contains new ingredients required for a precise lattice determination of higher moments of the PDFs and LCD.  First, as noted in Sec.~\ref{sec:convg}, we investigate carefully the convergence radius of the HOPE.   The analysis shown there also advances our understanding of the HOPE strategy, by indicating that large hadron momentum is required for the extraction of these higher moments.  Secondly, in this paper we report calculations for the one-loop Wilson coefficients which are required for a precise determination of the moments of the  unpolarised and helicity PDFs and meson LCDAs.  The Wilson coefficients for the PDFs can be extracted using the existing results~\cite{Gottschalk:1980rv,Gluck:1996ve} of cross sections of heavy-qaurk production in DIS involving flavour-changing currents.  Nevertheless, we perform independent calculations, as described in Secs.~\ref{sec:nlopdf} and \ref{sec:helicity}, and confirm these previous computations.  We emphasize that the one-loop Wilson coefficients for the LCDAs, presented in Sec.\ref{sec:danlo}, have not been previously calculated in the literature, and are thus new to this work.  

\section*{Acknowledgements}

CJL and RJP were supported by the Taiwanese MoST Grant No. 109-2112-M-009-006-MY3 and MoST Grant No. 109-2811-M-009-516.  The work of IK is partially supported by the MEXT as ``Program for Promoting Researches on the Supercomputer Fugaku'' (Simulation for basic science: from fundamental laws of particles to creation of nuclei) and JICFuS. YZ is supported in part by the U.S. Department of Energy, Office of Science, Office of Nuclear Physics, under grant Contract No. DE-SC0012704, No.~DE-AC02-06CH11357 and within the framework of the TMD Topical Collaboration. YZ is also supported by the U.S. Department of Energy, Office of Science, Office of Nuclear Physics and Office of Advanced Scientific Computing Research within the framework of Scientific Discovery through Advance Computing (ScIDAC) award Computing the Properties of Matter with Leadership Computing Resources. WD and AVG acknowledge support from the U.S.~Department of Energy (DOE) grant DE-SC0011090. WD is also supported within the framework of the TMD Topical Collaboration of the U.S.~DOE Office of Nuclear Physics, and by the SciDAC4 award DE-SC0018121.

\appendix

\section{Details of One-Loop Calculations}
\label{app:appendix}

In this appendix we provide the one-loop diagram-by-diagram results of the hadronic amplitude for the unpolarized PDF, helicity PDF and pseudoscalar meson LCDA cases. We only show the explicit expressions of the UV and IR divergent parts, and the expressions for finite parts can be found in the supplementary Mathematica notebook file.

\subsection{Unpolarized PDF}
\label{app:pdf}

The one-loop hadronic amplitude can be expressed as
\begin{align}\label{eq:upnloapp}
T_q^{\{\mu\nu\}(1)}\Big/ {\alpha_SC_F\over 4\pi}
	&={4i \bar{u}(p)q^{\{\mu} \gamma^{\nu\}} u(p)\over \tilde Q^2} H_{1}^{(1)} + {4 i \bar{u}(p)p^{\{\mu} \gamma^{\nu\}}u(p)\over \tilde Q^2} H_{2}^{(1)}\nn\\
	&\qquad-{ 2ig^{\mu\nu}_\perp  \bar{u}(p) \slashed q u(p)\over \tilde Q^2} H_{3}^{(1)} + {2i q^\mu q^\nu  \bar{u}(p) \slashed q u(p)\over \tilde Q^2} H_{4}^{(1)}\,.
\end{align}

The diagrammatic contributions are shown as follows:

The box diagram in \fig{box}.
\begin{align}
 H^{(1)}_{1, \rm box}(\tilde Q^2, \mu^2, \tau,\tilde \omega,\mu,\eps)&= \left({1\over \eps'_{\rm ir}} + \ln{\mu^2\over \tilde Q^2}\right)  {-2\tilde \omega - (1-\tilde \omega) \ln(1-\tilde \omega) + (1+\tilde \omega) \ln(1+\tilde \omega)\over \tilde \omega^2} + h^{(1)}_{1, \rm box}(\tau,\tilde \omega)\,,
\end{align}

\begin{align}
	H^{(1)}_{2, \rm box}(\tilde Q^2, \mu^2, \tau,\tilde \omega,\mu,\eps) 
	&= \left({1\over \eps'_{\rm ir}} + \ln{\mu^2\over \tilde Q^2}\right)  {-2\tilde \omega - (1-\tilde \omega) \ln(1-\tilde \omega) + (1+\tilde \omega) \ln(1+\tilde \omega)\over \tilde \omega^3} + h^{(1)}_{2, \rm box}(\tau,\tilde \omega)\,,
\end{align}

\begin{align}
	H^{(1)}_{3, \rm box}(\tilde Q^2, \mu^2, \tau,\tilde \omega,\mu,\eps) 
	&= \left({1\over \eps'_{\rm ir}} + \ln{\mu^2\over \tilde Q^2}\right)  {-2\tilde \omega - (1-\tilde \omega) \ln(1-\tilde \omega) + (1+\tilde \omega) \ln(1+\tilde \omega)\over \tilde \omega^2} + h^{(1)}_{3, \rm box}(\tau,\tilde \omega)\,,
\end{align}

\begin{align}
	H^{(1)}_{4, \rm box}(\tilde Q^2, \mu^2, \tau,\tilde \omega,\mu,\eps)
	&= h^{(1)}_{4, \rm box}(\tau,\tilde \omega)\,.
\end{align}

As one can see, the box diagram is UV finite but collinearly divergent.

The vertex diagrams in \figs{v1}{v2}.
\begin{align}
	H^{(1)}_{1, \rm vertex}(\tilde Q^2, \mu^2, \tau,\tilde \omega,\eps)
	&= \left({1\over \eps'_{\rm uv}} + \ln{\mu^2\over \tilde Q^2}\right) {2\tilde \omega\over 1- \tilde \omega^2}+ h^{(1)}_{1, \rm vertex}(\tau,\tilde \omega) \nn\\
	&\qquad -\left({1\over \eps'_{\rm ir}} + \ln{\mu^2\over \tilde Q^2}\right) \frac{2 \left[2 \tilde \omega ^2+\ln (1-\tilde \omega^2)-2 \tilde \omega  \tanh ^{-1}(\tilde \omega )\right]}{\tilde \omega \left(1- \tilde \omega ^2\right)}\,,
\end{align}

\begin{align}
	H^{(1)}_{2, \rm vertex}(\tilde Q^2, \mu^2, \tau,\tilde \omega,\eps) &= \left({1\over \eps'_{\rm uv}} + \ln{\mu^2\over \tilde Q^2}\right) {2\over 1- \tilde \omega^2}+ h^{(1)}_{2, \rm vertex}(\tau,\tilde \omega)\nn\\	
	&\qquad -\left({1\over \eps'_{\rm ir}} + \ln{\mu^2\over \tilde Q^2}\right) \frac{2 \left[2 \tilde \omega ^2+\ln (1-\tilde \omega^2)-2 \tilde \omega  \tanh ^{-1}(\tilde \omega )\right]}{\tilde \omega ^2 \left(1- \tilde \omega ^2\right)}\,,
\end{align}

\begin{align}
	H^{(1)}_{3, \rm vertex}(\tilde Q^2, \mu^2, \tau,\tilde \omega,\eps)&= \left({1\over \eps'_{\rm uv}} + \ln{\mu^2\over \tilde Q^2}\right) {2\tilde \omega\over 1- \tilde \omega^2}+ h^{(1)}_{3, \rm vertex}(\tau,\tilde \omega)\nn\\	
	&\qquad -\left({1\over \eps'_{\rm ir}} + \ln{\mu^2\over \tilde Q^2}\right) \frac{2 \left[2 \tilde \omega ^2+\ln (1-\tilde \omega^2)-2 \tilde \omega  \tanh ^{-1}(\tilde \omega )\right]}{\tilde \omega \left(1- \tilde \omega ^2\right)}\,,
\end{align}

\begin{align}
	H^{(1)}_{4, \rm vertex}(\tilde Q^2, \mu^2, \tau,\tilde \omega,\eps)	&= 0\,.
\end{align}

The self-energy diagrams in \fig{self} and \fig{self2}.
\begin{align}
	H^{(1)}_{1, \rm self}(\tilde Q^2, \mu^2, \tau,\tilde \omega,\eps) &=- \left({1\over \eps'_{\rm uv}} - {1\over \eps'_{\rm ir}}\right) {\tilde \omega \over 1-\tilde \omega^2} - \left({1\over \eps'_{\rm uv}} + \ln{\mu^2\over \tilde Q^2}\right) {\tilde \omega \over 1- \tilde \omega^2} - (1-\tau)\left({3\over \eps'_{\rm uv}} + 3\ln{\mu^2\over \tilde Q^2}\right) {4\tilde \omega \over (1-\tilde \omega^2)^2} \nn\\
	& \qquad + h^{(1)}_{1, \rm self}(\tau,\tilde \omega)\,,
\end{align}
where the first term comes from the light-quark self-energy, while the second and third terms originate fom the heavy-quark self-energy. 

\begin{align}
	H^{(1)}_{2, \rm self}(\tilde Q^2, \mu^2, \tau,\tilde \omega,\eps)&=- \left({1\over \eps'_{\rm uv}} - {1\over \eps'_{\rm ir}}\right) {1\over 1-\tilde \omega^2} - \left({1\over \eps'_{\rm uv}} + \ln{\mu^2\over \tilde Q^2}\right) {1\over 1- \tilde \omega^2} - (1-\tau)\left({3\over \eps'_{\rm uv}} + 3\ln{\mu^2\over \tilde Q^2}\right) {2(1+\tilde \omega^2)\over (1-\tilde \omega^2)^2}\nn\\
	&\qquad  + h^{(1)}_{2, \rm self}(\tau,\tilde \omega)\,,
\end{align}

\begin{align}
	H^{(1)}_{3, \rm self}(\tilde Q^2, \mu^2, \tau,\tilde \omega,\eps) &=- \left({1\over \eps'_{\rm uv}} - {1\over \eps'_{\rm ir}}\right) {\tilde \omega \over 1-\tilde \omega^2} - \left({1\over \eps'_{\rm uv}} + \ln{\mu^2\over \tilde Q^2}\right) {\tilde \omega \over 1- \tilde \omega^2} - (1-\tau)\left({3\over \eps'_{\rm uv}} + 3\ln{\mu^2\over \tilde Q^2}\right) {4\tilde \omega \over (1-\tilde \omega^2)^2} \nn\\
	& \qquad + h^{(1)}_{3, \rm self}(\tau,\tilde \omega)\,,
\end{align}

\begin{align}
	H^{(1)}_{4, \rm self}(\tilde Q^2, \mu^2, \tau,\tilde \omega,\eps) &= 0 \,.
\end{align}

The heavy-quark mass needs renormalization, and the corresponding counter-terms in the $\MS$ and on-shell schemes are
\begin{align}
	\delta H_{1,m}^{\MS(1)} &= \delta H_{3,m}^{\MS(1)} = (1-\tau){3\over \eps'_{\rm uv}} {4\tilde \omega \over (1-\tilde \omega^2)^2} \,,\\
	\delta H_{1,m}^{\rm OS(1)} &= \delta H_{3,m}^{\rm OS(1)} = (1-\tau)\left[{3\over \eps'_{\rm uv}} +4 + 3\ln{\mu^2\over m_\Psi^2}\right] {4\tilde \omega \over (1-\tilde \omega^2)^2}\,.
\end{align}
\begin{align}
	\delta H_{2,m}^{\MS(1)} &= (1-\tau){3\over \eps'_{\rm uv}} {2(1+\tilde \omega^2)\over (1-\tilde \omega^2)^2} \,,\\
	\delta H_{2,m}^{\rm OS(1)} &= (1-\tau)\left[{3\over \eps'_{\rm uv}} +4 + 3\ln{\mu^2\over m_\Psi^2}\right] {2(1+\tilde \omega^2)\over (1-\tilde \omega^2)^2}\,.
\end{align}

By summing all the one-loop diagrams, one finds out that the UV divergences regulated by $\eps'_{\rm uv}$ cancel except for that from the heavy-quark mass correction. The collinear divergences regulated by $\eps'$ add up to
\begin{align}
- {1\over\eps'}\frac{\tilde \omega  \left(2+\tilde \omega ^2\right)+\left(1+\tilde \omega ^2\right) \left[\tilde \omega  \ln{(1-\tilde \omega^2)}-2 \tanh ^{-1}(\tilde \omega )\right]}{\tilde \omega ^2 \left(1-\tilde \omega ^2\right)}
\end{align}
in $H^{(1)}_{1}$ and $H^{(1)}_{3}$ and 
\begin{align}
	- {1\over\eps'}\frac{\tilde \omega  \left(2+\tilde \omega ^2\right)+\left(1+\tilde \omega ^2\right) \left[\tilde \omega  \ln{(1-\tilde \omega^2)}-2 \tanh ^{-1}(\tilde \omega )\right]}{\tilde \omega ^3 \left(1-\tilde \omega ^2\right)}
\end{align}
in $H^{(1)}_{2}$, whose Taylor expansions in $\tilde \omega$ exactly reproduce the collinear divergences in the Mellin moments in the OPE in \eq{tnlo}. This is an important consistency check of our results.

In the limit of massless heavy-quark, i.e., $\tau\to1$, the hadronic amplitude no longer depends on the heavy-quark mass renormalization. Moreover, it satisfies the chiral ward identity
\begin{align}
	q_\mu T_q^{\mu\nu}(q,p,m_\Psi=0,\mu,\epsilon) = q_\nu T_q^{\mu\nu}(q,p,m_\Psi=0,\mu,\epsilon) = 0\,,
\end{align}
which leads to the relations
\begin{align}
	H_1 &=\tilde \omega  H_2\,,\ \ \ \ \ \ \ \ H_1 = H_3+H_4\,,
\end{align}
as we have also checked.

All the finite parts, $h^{(1)}_{i, \rm box}$, $h^{(1)}_{i, \rm vertex}$ and $h^{(1)}_{i, \rm self}$ are provided in the supplementary material.  Therefore, the finite non-logarithmic part of the hadronic amplitudes in the $\MS$ scheme are
\begin{align}
	h^{\MS (1)}_1(\tilde Q^2, \mu^2, \tau, \tilde \omega) &= h^{(1)}_{1,\rm box}(\tau,\tilde \omega)+h^{(1)}_{1,\rm vertex}(\tau,\tilde \omega)+h^{(1)}_{1,\rm self}(\tau,\tilde \omega)  { - (1-\tau)\left(3\ln{\mu^2\over \tilde Q^2}\right) {4\tilde \omega \over (1-\tilde \omega^2)^2}}\,, \nn\\
	h^{\MS (1)}_2(\tilde Q^2, \mu^2, \tau, \tilde \omega) &= h^{(1)}_{2,\rm box}(\tau,\tilde \omega)+h^{(1)}_{2,\rm vertex}(\tau,\tilde \omega)+h^{(1)}_{2,\rm self}(\tau,\tilde \omega)  { - (1-\tau)\left(3\ln{\mu^2\over \tilde Q^2}\right) {2(1+\tilde \omega^2)\over (1-\tilde \omega^2)^2}}\,, \nn\\
	h^{\MS (1)}_3(\tilde Q^2, \mu^2, \tau, \tilde \omega) &= h^{(1)}_{3,\rm box}(\tau,\tilde \omega)+h^{(1)}_{3,\rm vertex}(\tau,\tilde \omega)+h^{(1)}_{3,\rm self}(\tau,\tilde \omega)  { - (1-\tau)\left(3\ln{\mu^2\over \tilde Q^2}\right) {4\tilde \omega \over (1-\tilde \omega^2)^2}}\,, \nn\\
	h^{\MS (1)}_4(\tau, \tilde \omega) &= h^{(1)}_{4,\rm box}(\tau,\tilde \omega)\,,
\end{align}
whose Taylor expansion in $\tilde \omega$ gives the $\MS$ Wilson coefficients. The on-shell scheme result is related to $h^{\MS (1)}$ by
\begin{align}
	h^{{\rm OS} (1)}_1(\tau, \tilde \omega) - h^{\MS (1)}_1(\tilde Q^2, \mu^2, \tau, \tilde \omega) &= h^{{\rm OS} (1)}_3(\tau, \tilde \omega) - h^{\MS (1)}_3(\tilde Q^2, \mu^2, \tau, \tilde \omega) = (1-\tau){4\tilde \omega \over (1-\tilde \omega^2)^2} \Big(4 { + 3\ln{\mu^2\over m_\Psi^2}}\Big) \,.\\
	h^{{\rm OS} (1)}_2(\tau, \tilde \omega) - h^{\MS (1)}_2(\tilde Q^2, \mu^2, \tau, \tilde \omega) &= (1-\tau){2(1+\tilde \omega^2)\over (1-\tilde \omega^2)^2} \Big(4 {+ 3\ln{\mu^2\over m_\Psi^2}}\Big) \,,\\
	h^{{\rm OS} (1)}_4(\tau, \tilde \omega) - h^{\MS (1)}_4(\tilde Q^2, \mu^2, \tau, \tilde \omega) &= 0  \,.
\end{align}

\subsection{Helicity PDF}
\label{app:hpdf}

The one-loop correction to the hadronic amplitude for the helicity PDF case can be expressed as
\begin{align}
	T_q^{[\mu\nu](1)}&= -{2\epsilon^{\mu\nu\rho\sigma}q_\rho  \over \tilde Q^2} \ \bar{u}(p)\gamma_\sigma \gamma_5 u(p)\ {\alpha_SC_F\over 4\pi}H^{(1)}\,.
\end{align}
To simplify the calculation, we can choose $p^\mu=(p^z,0,0,p^z)$ and $\mu, \nu = 1, 2$ without loss of generality.

The diagrammatic contributions are as follows:

The box diagram in \fig{box}.
\begin{align}
	H^{(1)}_{\rm box}(\tilde Q^2, \mu^2, \tau,\tilde \omega,\eps)
	&= - \left({1\over \eps'_{\rm ir}} + \ln{\mu^2\over \tilde Q^2}\right) {(1-\tilde \omega) \ln(1-\tilde \omega) + (1+\tilde \omega) \ln(1+\tilde \omega)\over \tilde \omega^2} + h^{(1)}_{\rm box}(\tau,\tilde \omega)\,,
\end{align}
which is UV finite but collinearly divergent.

Note that since the helicity PDF is defined with the axial vector current, we must treat the definition of $\gamma_5$ with care. In our calculation, we adopt the scheme used in Refs.~\cite{delAguila:1981nk,Braaten:1982yp}, where the treatment of $\gamma_5$ is determined by calibrating the Wilson coefficients to those obtained with quark off-shellness $p^2<0$ as the regulator for collinear divergence. Within this scheme, the moments of the helicity PDF are the same as the unpolarized case in \eq{moment1}, while only the box diagram must be modified, as one has to introduce a finite counter-term that is equivalent to the replacement,
\begin{align}\label{eq:gamma5}
	{1\over \eps'_{\rm ir}} \to {1\over \eps'_{\rm ir}}  - 4\,.
\end{align}
This finite counter-term induced by the factor $(-4)$ has been absorbed into $h^{(1)}_{\rm box}(\tau,\tilde \omega)$.

The vertex diagrams in \figs{v1}{v2}.
\begin{align}
	H^{(1)}_{\rm vertex}(\tilde Q^2, \mu^2, \tau,\tilde \omega,\eps) 
	&= \left({1\over \eps'_{\rm uv}} + \ln{\mu^2\over \tilde Q^2}\right) {2\over 1- \tilde \omega^2} + h^{(1)}_{\rm vertex}(\tau,\tilde \omega)\nn\\
	&\qquad  \!-\! \left({1\over \eps'_{\rm ir}} + \ln{\mu^2\over \tilde Q^2}\right) \left[{4\over 1-\tilde \omega^2} \!+\! {2\over \tilde \omega}\left({\ln(1-\tilde \omega)\over 1-\tilde \omega} \!-\! {\ln(1+\tilde \omega)\over 1+\tilde \omega}\right)\right] \,.
\end{align}

The self-energy diagrams in \fig{self} and \fig{self2}.
\begin{align}
	H^{(1)}_{\rm self}(\tilde Q^2, \mu^2, \tau,\tilde \omega,\eps)
	&=- \left({1\over \eps'_{\rm uv}} - {1\over \eps'_{\rm ir}}\right) {1\over 1-\tilde \omega^2} - \left({1\over \eps'_{\rm uv}} + \ln{\mu^2\over \tilde Q^2}\right) {1\over 1- \tilde \omega^2}   - (1-\tau)\left({3\over \eps'_{\rm uv}} + 3\ln{\mu^2\over \tilde Q^2}\right) {2(1+\tilde \omega^2)\over (1-\tilde \omega^2)^2} \nn\\
	& \qquad + h^{(1)}_{\rm self}(\tau,\tilde \omega)\,.
\end{align}

The heavy-quark mass needs renormalization, and the corresponding counter-terms in the $\MS$ and on-shell schemes are
\begin{align}
	\delta H_m^{\MS(1)} &= (1-\tau){3\over \eps'_{\rm uv}} {2(1+\tilde \omega^2)\over (1-\tilde \omega^2)^2} \,,\\
	\delta H_m^{\rm OS(1)} &= (1-\tau)\left[{3\over \eps'_{\rm uv}} +4 + 3\ln{\mu^2\over m_\Psi^2}\right] {2(1+\tilde \omega^2)\over (1-\tilde \omega^2)^2}\,.
\end{align}

Again, one finds that the UV divergences regulated by $\eps'_{\rm uv}$ cancel except for that from the heavy-quark mass correction. Besides, collinear divergences regulated by $\eps'$ add up to
\begin{align}
	-{1\over\eps'}\left[{3\over 1-\tilde \omega^2} + {1+\tilde \omega^2\over \tilde \omega^2}\left({\ln(1-\tilde \omega)\over 1-\tilde \omega} + {\ln(1+\tilde \omega)\over 1+\tilde \omega}\right)  \right]\,,
\end{align}
whose Taylor expansion in $\tilde \omega$ exactly reproduces the collinear divergences of the Mellin moments in \eq{tnlo2}.

All the finite parts, $h^{(1)}_{\rm box}$, $h^{(1)}_{\rm vertex}$ and $h^{(1)}_{\rm self}$ are provided in the supplementary material. Therefore, the finite non-logarithmic part of the hadonic amplitude is
\begin{align}
	h^{\MS (1)}(\tilde Q^2, \mu^2, \tau, \tilde \omega) &= h^{(1)}_{\rm box}(\tau,\tilde \omega)+h^{(1)}_{\rm vertex}(\tau,\tilde \omega)+h^{(1)}_{\rm self}(\tau,\tilde \omega) {  - (1-\tau)\left( 3\ln{\mu^2\over \tilde Q^2}\right) {2(1+\tilde \omega^2)\over (1-\tilde \omega^2)^2}}
\end{align}
in the $\MS$ scheme, whose Taylor expansion in $\tilde \omega$ gives the $\MS$ Wilson coefficients. The on-shell scheme result is related to $h^{\MS (1)}$ by
\begin{align}
	h^{{\rm OS} (1)}(\tau, \tilde \omega) - h^{\MS (1)}(\tilde Q^2, \mu^2, \tau, \tilde \omega) &= (1-\tau){2(1+\tilde \omega^2)\over (1-\tilde \omega^2)^2} \Big(4 { + 3\ln{\mu^2\over m_\Psi^2}}\Big) \,.
\end{align}

\subsection{LCDA}
\label{app:lcda}

The one-loop correction to the hadronic amplitude for the LCDA case can be expressed as
\begin{align}
	V^{[\mu\nu](1)}&= -{\alpha_sC_F\over 4\pi} {2 \epsilon^{\mu\nu\rho\sigma}q_\rho \over \tilde{Q}^2}\bar{v}\gamma_\sigma \gamma_5 u\ H^{(1)}\,.
\end{align}
To simplify the calculation, we can choose $p^\mu=(p^z,0,0,p^z)$, $q^\mu=(q^0,0,0,q^z)$ and $\mu, \nu = 1, 2$ without loss of generality.

The one-loop contributions are as follows:

The box diagram in \fig{box}.
\begin{align}
	H^{(1)}_{\rm box}(\tilde Q^2, \mu^2, \tau,\tilde \omega, x_0,\eps)
	&= - \left({1\over \eps'_{\rm ir}} + \ln{\mu^2\over \tilde Q^2}\right) {4\over (1-x_0^2) \tilde \omega^2}\left[\ln \frac{4-\tilde \omega^2}{4-x^2_0 \tilde \omega^2} \!-\! x_0 \tilde \omega  \tanh ^{-1}\left(\frac{x_0 \tilde \omega }{2}\right) \!+\!\tilde \omega  \tanh ^{-1}\left(\frac{\tilde \omega }{2}\right)\right] \nn\\
	&\qquad + h^{(1)}_{\rm box}(\tau,\tilde \omega,x_0)\,,
\end{align}
where we treat $\gamma_5$ with the same prescription for the helicity PDF case by making the substitution in \eq{gamma5}.

The vertex diagrams in \figs{v1}{v2}.
\begin{align}
	H^{(1)}_{\rm vertex}(\tilde Q^2, \mu^2, \tau,\tilde \omega, x_0,\eps)
	&= \left({1\over \eps'_{\rm uv}} + \ln{\mu^2\over \tilde Q^2}\right) {8\over 4- x_0^2\tilde \omega^2}+ h^{(1)}_{\rm vertex}(\tau,\tilde \omega,x_0)\\	
	&\quad + \left({1\over \eps'_{\rm ir}} + \ln{\mu^2\over \tilde Q^2}\right) {8\over (1-x^2_0) \tilde \omega (4-x_0^2 \tilde \omega^2)}\nn\\
	&\quad \times \left[-(1-x^2_0) \tilde \omega  \left(\ln \frac{4-x^2_0 \tilde \omega ^2}{4-\tilde \omega ^2}+2\right)  +\left(4\!-\!x^2_0 \tilde \omega ^2\right) \tanh ^{-1}\left(\frac{\tilde \omega }{2}\right) \!-\! x_0 (4\!-\!\tilde \omega ^2) \tanh ^{-1}\left(\frac{x_0 \tilde \omega }{2}\right)\right]\,.\nn
\end{align}

The self-energy diagrams in \fig{self} and \fig{self2}.
\begin{align}
	H^{(1)}_{\rm self}(\tilde Q^2, \mu^2, \tau,\tilde \omega,x_0,\eps)
	&=- \left({1\over \eps'_{\rm uv}} - {1\over \eps'_{\rm ir}}\right) {4\over 4- x_0^2\tilde \omega^2} - \left({1\over \eps'_{\rm uv}} + \ln{\mu^2\over \tilde Q^2}\right) {4\over 4- x_0^2\tilde \omega^2}  - (1-\tau)\left({3\over \eps'_{\rm uv}} + 3\ln{\mu^2\over \tilde Q^2}\right) {8(4+x_0^2\tilde \omega^2)\over (4-x_0^2\tilde \omega^2)^2}  \nn\\
	& \qquad + h^{(1)}_{\rm self}(\tau,\tilde \omega,x_0)\,.
\end{align}

The heavy-quark mass needs renormalization, and the corresponding counter-terms in the $\MS$ and on-shell schemes are
\begin{align}
	\delta H_m^{\MS(1)} &= (1-\tau){3\over \eps'_{\rm uv}}{8(4+x_0^2\tilde \omega^2)\over (4-x_0^2\tilde \omega^2)^2} \,,\\
	\delta H_m^{\rm OS(1)} &= (1-\tau)\left[{3\over \eps'_{\rm uv}} +4 + 3\ln{\mu^2\over m_\Psi^2}\right]{8(4+x_0^2\tilde \omega^2)\over (4-x_0^2\tilde \omega^2)^2}\,.
\end{align}

Same as the PDF cases, the UV divergences regulated by $\eps'_{\rm uv}$ cancel except for that from the heavy-quark mass correction. The collinear divergences add up to \eq{lcdall}, which also passes the consistency check that its conformal expansion reproduces the collinear divergences of the Gegenbauer moments.

All the finite parts, $h^{(1)}_{\rm box}$, $h^{(1)}_{\rm vertex}$ and $h^{(1)}_{\rm self}$, as well as their massless limits when $\tau\to1$, are provided in the supplementary material. We can identify that the finite term ${\mathbf R}^{(1)}(\tau, \tilde \omega, x_0)$ in \eq{oneloop} is
\begin{align}
	{\mathbf R}^{\MS (1)}(\tilde Q^2, \mu^2, \tau, \tilde \omega, x_0) &= h^{(1)}_{\rm box}(\tau,\tilde \omega,x_0)+h^{(1)}_{\rm vertex}(\tau,\tilde \omega,x_0)+h^{(1)}_{\rm self}(\tau,\tilde \omega,x_0) { - (1-\tau)\left( 3\ln{\mu^2\over \tilde Q^2}\right) {8(4+x_0^2\tilde \omega^2)\over (4-x_0^2\tilde \omega^2)^2}}
\end{align}
in the $\MS$ scheme. The on-shell scheme result is related to ${\mathbf R}^{\MS (1)}$ by
\begin{align}
	{\mathbf R}^{{\rm OS} (1)}(\tau, \tilde \omega, x_0) - {\mathbf R}^{\MS (1)}(\tilde Q^2, \mu^2, \tau, \tilde \omega, x_0) &= (1-\tau){8(4+x_0^2\tilde \omega^2)\over (4-x_0^2\tilde \omega^2)^2} \Big(4 {+3\ln{\mu^2\over m_\Psi^2}}\Big) \,.
\end{align}

Notably, in the massless limit, our result of ${\mathbf R}^{(1)}(1, \tilde \omega, x_0)$
reproduces Eq.~(5.2) in the literature~\cite{Braaten:1982yp} if make the replacements 
\begin{align}
	Q^2\to 2Q^2\,,\ \ \ \ \ x\to {1\over 2}(1-x_0)\,,\ \ \ \ \  w\to {1\over2} - {\omega\over4}.
\end{align}

\bibliography{bibliography}

\end{document}